# Mantodea phylogenomics provides new insights into X-chromosome progression and evolutionary radiation

Hangwei Liu[1,2,†], Lihong Lei[1,3,4,†], Fan Jiang[1], Bo Zhang[1], Hengchao Wang[1], Yutong Zhang[3], Anqi Wang[1], Hanbo Zhao[1], Guirong Wang[1,5,*] & Wei Fan[1,*]

[1] Guangdong Laboratory for Lingnan Modern Agriculture (Shenzhen Branch), Genome Analysis Laboratory of the Ministry of Agriculture and Rural Affairs, Agricultural Genomics Institute at Shenzhen, Chinese Academy of Agricultural Sciences, Shenzhen, Guangdong, 518120, China.

[2] College of Plant Protection, Yangzhou University, Yangzhou 225009, China

[3] School of Life Sciences, Henan University, Kaifeng 475004, China

[4] Shenzhen Research Institute of Henan University, Shenzhen 518000, China

[5] State Key Laboratory for Biology of Plant Diseases and Insect Pests, Institute of Plant Protection, Chinese Academy of Agricultural Sciences, Beijing, China.

†These authors contributed equally to this work.

* Correspondence should be addressed to wangguirong@caas.cn and fanwei@caas.cn.

## Abstract

**Background** Praying mantises, members of the order Mantodea, play important roles in agriculture, medicine, bionics, and entertainment. However, the scarcity of genomic resources has hindered extensive studies on mantis evolution and behaviour.

**Results** Here, we present the chromosome-scale reference genomes of five mantis species: the European mantis (*Mantis religiosa*), Chinese mantis (*Tenodera sinensis*), triangle dead leaf mantis (*Deroplatys truncata*), orchid mantis (*Hymenopus coronatus*), and metallic mantis (*Metallyticus violaceus*). We found that transposable element expansion is the major force governing genome size in Mantodea. Based on whole-alignments, we deduced that the Mantodea ancestor may have had only one X chromosome and that translocations between the X chromosome and an autosome may have occurred in the lineage of the superfamily Mantoidea. Furthermore, we found a lower evolutionary rate for the metallic mantis than for the other mantises. We also found that Mantodea underwent rapid radiation after the K-Pg mass extinction event, which could have contributed to the confusion in species classification.

**Conclusions** We present the chromosome-scale reference genomes of five mantis species to reveal the X-chromosome evolution, clarify the phylogeny relationship, and transposable element expansion.

**Keywords: Mantodea, genome, transposable elements,X- chromosome, radiation**

**Background**

Praying mantises are familiar insects that play important roles in agriculture, medicine, bionics, and entertainment. Mantodea has evolved into a group comprising ~2500 species with diverse morphological and ecological characteristics, with the highest diversity in tropical and subtropical habitats [1, 2]. As predators of many harmful insect species, praying mantises such as the European mantis (*Mantis religiosa*) and Chinese mantis (*Tenodera sinensis*) are widely acknowledged as natural enemies that control plant pests[3], benefiting organic planting where pesticide is prohibited. Praying mantises also exhibit many unique and interesting phenotypes, which are essential for understanding phenotypic innovations. The mantis ootheca (egg capsule, egg chamber) is a traditional medicine used to cure frequent micturition, strengthen kidney health and prevent spermatorrhea in East Asian countries [4]. Most praying mantises have two sharp and strong forelegs, which are much larger and more powerful than their ancient ancestors. In addition, the femur and tibia of the forelegs are armed with strong spines along their posterior edges. When the femur and tibia fold on each other, a praying mantis can firmly grasp the prey [4]. This distinctive body structure of the praying mantis has been a significant source of inspiration in bionics. For example, a more efficient bionic cutting blade was designed based on the forelegs of mantises [5, 6]. Finally, several species, such as the triangle dead leaf mantis (*Deroplatys truncata*) and orchid mantis (*Hymenopus coronatus*), have important skills that help in capturing prey along with camouflage, a key adaptive strategy for avoiding predators or attracting prey [7].

The two closely related orders, Mantodea (mantises) and Blattodea (cockroaches and termites), are classified into the superorder Dictyoptera, and phylogenomic analyses revealed that Mantodea split from Blattodea during the Permian [8]. From fossils of early Mantodea and Blattodea species, the common ancestor is thought to resemble modern cockroaches in many aspects [9]. The metallic mantis (*Metallyticus violaceus*) has many morphological features similar to those of modern cockroaches, suggesting that the common ancestor of Dictyoptera has a cockroach-like morphology [10]. Although Mantodea are well supported as monophyletic, the phylogenetic relationships within Mantodea are still not well resolved. Morphological and molecular phylogenies are often inconsistent with previous classification schemes, because of the rapid radiation events and convergent evolution of ecomorphological strategies [1, 11].

Compared with those of many other insect orders, the genomic resources of Mantodea are very limited, with only three chromosome-scale reference genomes available: the Chinese mantis (*T. sinensis*), orchid mantis (*H. coronatus*) and Malaysian dead leaf mantis (*Deroplatys lobata*) [12, 13]. Here, we present the chromosome-scale reference genomes of three other mantis species, the European mantis (*M. religiosa*), triangle dead leaf mantis (*D. truncata*), and metallic mantis (*M. violaceus*), as well as a more complete assembly of *T. sinensis* and *H. coronatus*, to reveal X chromosome evolution and clarify phylogenetic relationships.

**Results**

**Chromosome-scale genome assemblies of five mantis species**

We generated 179 Gb (48X) (*M. religiosa*), 97 Gb (36X) (*T. sinensis*), 112 Gb (26X) (*D. truncata*), 177 Gb (56X) (*H. coronatus*), and 147 Gb (63X) (*M. violaceus*) PacBio HiFi data, and 127 Gb (35X) (*M. religiosa*), 112 Gb (42X) (*T. sinensis*), 153 Gb (35X) (*D. truncata*), 182 Gb (58X) (*H. coronatus*), and 183 Gb (79X) (*M. violaceus*) Illumina Hi-C data (Table S1, S2). The PacBio HiFi reads were used to assemble the contig sequences, with a total size of 3.6 Gb (*M. religiosa*), 2.6 Gb (*T. sinensis*), 4.2 Gb (*D. truncata*), 3.1 Gb (*H. coronatus*), and 2.3 Gb (*M. violaceus*) and N50 sizes of 1 Mb (*M. religiosa*), 13 Mb (*T. sinensis*), 44 Mb (*D. truncata*), 71 Mb (*H. coronatus*), and 109 Mb (*M. violaceus*). The Illumina Hi-C reads were mapped to the contig sequences, and the valid Hi-C read pairs were used for scaffolding assembly (Table S3), resulting in 85.39% (*M. religiosa*), 95.63% (*T. sinensis*), 97.47% (*D. truncata*), 98.27% (*H. coronatus*), and 98.51% (*M. violaceus*) of the contig sequences anchored into 14 (*M. religiosa*), 14 (*T. sinensis*), 16 (*D. truncata*), 21 (*H. coronatus*), and 17 (*M. violaceus*) chromosome-level scaffolds (Figure 1A-E, S1, S2; Table 1, S4). Notably, only the chromosome numbers for European and Chinese mantises have been karyotyped [14, 15], whereas the others are inferred only from the genome assembly. In the GCE method [16], the estimated genome sizes are 3.5 Gb (*M. religiosa*), 2.8 Gb (*T. sinensis*), 4.3 Gb (*D. truncata*), 3.1 Gb (*H. coronatus*), and 2.3 Gb (*M. violaceus*), consistent with assembled genome sizes. Owing to the higher heterozygosity rate, the contig sizes for the European and Chinese mantids are shorter than those for the other three species (Figure S3).

**Table 1. Statistics of genome assembly and annotation**

| Genomic features | *M. religiosa* | *T. sinensis* | *D. truncata* | *H. coronatus* | *M. violaceus* |
|---|---|---|---|---|---|
| Genome assembly | | | | | |
| Estimated genome size by K-mer (bp) | 3,519,843,697 | 2,865,686,147 | 4,337,798,490 | 3,167,239,197 | 2,331,221,057 |
| Total assembly size (bp) | 3,680,002,721 | 2,687,426,722 | 4,290,792,545 | 3,127,590,514 | 2,322,129,794 |
| Contig N50 size (bp) | 1,407,320 | 12,728,340 | 44,444,664 | 71,519,735 | 109,157,195 |
| Scaffold N50 size (bp) | 210,326,877 | 190,002,057 | 248,405,437 | 159,059,693 | 125,733,329 |
| # of assembly-inferred chromosomes | 14 | 14 | 16 | 21 | 17 |
| % sequence anchored to chromosome | 85.39% | 95.63% | 97.47% | 98.27% | 98.51% |
| Genome annotation | | | | | |
| Length and % of tandem sequences (bp) | 396,842,330 (10.8%) | 403,304,947 (15.0%) | 471,243,565 (11.0%) | 238,530,960 (7.6%) | 186,949,249 (8.1%) |
| Length and % of TE sequences (bp) | 2,501,898,483 (68%) | 1,710,668,926 (64%) | 2,928,636,453 (68%) | 2,122,785,940 (68%) | 1,351,077,317 (58%) |
| Number of protein-coding gene models | 19,017 | 19,007 | 18,156 | 18,536 | 17,804 |
| Mean CDS length (bp) | 1551 | 1782 | 1601 | 1523 | 1152 |
| Mean exon number | 6.07 | 5.93 | 6.34 | 6.33 | 5.54 |

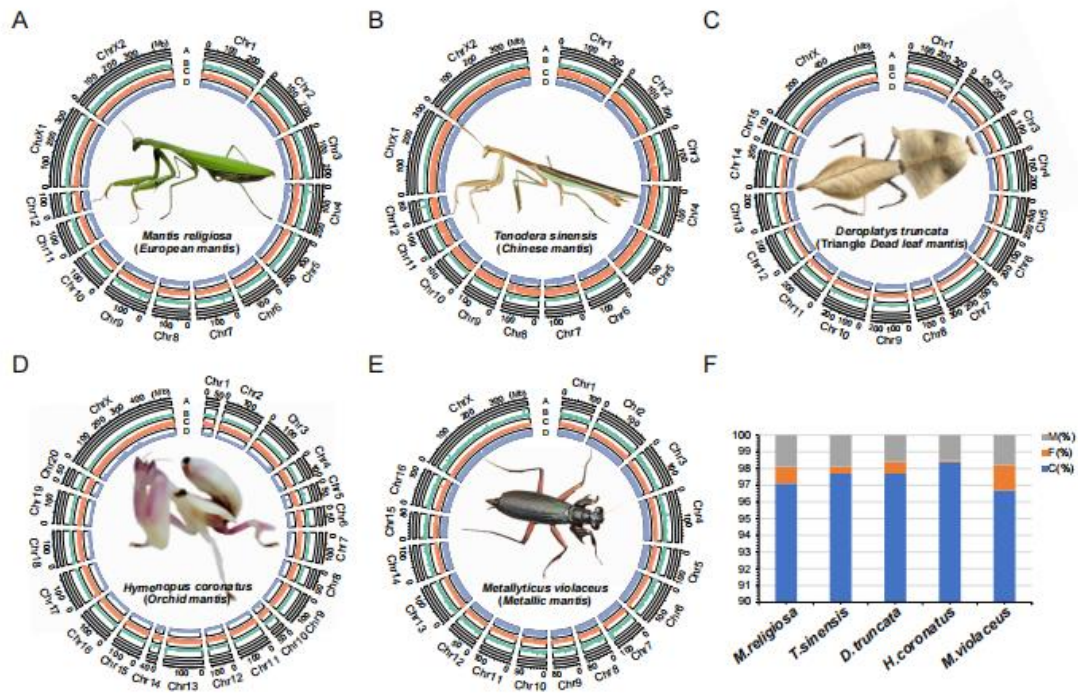

**Figure 1**. Circos plots of genomic annotations for *M. religiosa.* (A) *T. sinensis* (B) *D. truncata* (C) *H. coronatus* (D) and *M. violaceus* (E). Each circos plot has 4 tracks: track A represents chromosome length, track B represents gene density, track C represents transposable element (TE) density, and track D represents GC percentage. Feature density and GC percentage were calculated by sliding 1-Mb windows. (F) BUSCO assessment (database: Insecta from OrthoDB v10) of gene sets for five mantis species. M means missing, F means fragmented, and C means complete.

Recently, Huang et al. published a reference genome for the orchid mantis, with a much shorter contig N50 size of 15.7 Mb [12], and Yuan et al. published a reference genome of *T. sinensis* with a contig N50 size of 2.36 Mb, which is also much shorter than that of this study [13]. From syntenic alignments of the two assemblies for the orchid mantis, we found that most chromosomes were largely consistent except for the X chromosome (Figure S4). One complete X chromosome in our assembly corresponds to 3 fragmented chromosomes in Huang's assembly. The X chromosome is the largest chromosome, making it more difficult to assemble than the autosomes. Thus, our assembly of the X chromosome for the orchid mantis is likely more complete. We also compared another reference genome published by the Huang group [12], the Malaysian dead leaf mantis, to our assembled reference genome of the triangle dead leaf mantis (Figure S5). Most chromosomes belonging to the same genus, have high synteny, but four chromosomes are involved in chromosome-level rearrangements, which are more likely due to species divergence than assembly errors. Both reference genomes of the Chinese mantis (Yuan et al.[13] and the present study) showed high synteny for all chromosomes (Figure S6).

By integrating homology and transcription evidence, 19,017 (*M. religiosa*), 19,007 (*T. sinensis*), 18,156 (*D. truncata*), 18,536 (*H. coronatus*) and 17,804 (*M. violaceus*) protein-coding gene models were predicted (Table 1, Table S1, S7). The BUSCO complete rates for the reference gene sets of these mantis species range from 96.7%-98.4% (Figure 1F), which are higher than or comparable to those of previously published *Dictyoptera* genomes [12, 17, 18]. Furthermore, 97.2%-98.6% of the genes in these five mantis species were assigned functions according to at least one of the NCBI-NR, KEGG, InterPro or GO databases.

**TE expansions enlarge Mantodea genome and vary among lineages**
Increasing evidence has shown that transposable elements (TEs) contribute significantly to total genome size and influence genome architecture, along with insertions, deletions, translocations, etc [19]. Although most TE copies have no visible effect on fitness, some TE insertion events have been shown to significantly benefit host organisms, thus playing an important role in their evolution [20]. We analysed the total TE content among the 5 species and found that genome size was linearly correlated with TE abundance (Figure S7). Chinese and metallic mantises have relatively smaller genome sizes (2.3-2.8 Gb) and lower TE contents (58-63%), than the other 3 mantises, with relatively larger genome sizes (3.1-3.5 Gb) and higher TE contents (67-68%), suggesting that genome size differences are mostly determined by TE expansion in mantids.

Many retrotransposons, DNA transposons and rolling-circle transposons were found in these genomes; however, their ratios differ across species (Figure 2A). Among the two Mantidae species, LINEs are the largest components, and a sharp expansion of LINEs with divergence of approx. 7% was found in the European mantis (Figure 2B). However, no recent large-scale expansion of LINEs has occurred in the Chinese mantis, which may explain why its genome size (2.8 Gb) is smaller than that of the European mantis (3.5 Gb). Unlike the two Mantidae species, orchid and triangle dead leaf mantis have massive DNA transposons, with Tc1 (especially Tc1-IS630-Pogo) being the largest component in these two species, which is consistent with the findings of a former study [12]. The triangle dead leaf mantis has undergone both a recent sharp expansion and an ancient burst of Tc1 in its genome, leading to the largest genome size (4.3 Gb) found in this study. Only an ancient explosion of Tc1 was observed in the orchid mantis (Figure 2C). Both orchid and triangle dead leaf mantises also have a large rolling-circle transposon, *Helitrons*. Both a recent and an ancient burst of *Helitron* were observed in the triangle dead leaf mantis, whereas only an ancient burst of *Helitron* was found in the orchid mantis (Figure 2D). The metallic mantis showed no recent accumulation of any category of TEs, which may explain why its genome size (2.3 Gb) was the smallest among the mantises. These results collectively suggest that TE expansion is the major force behind genome size variation in Mantodea, whose genome sizes are larger than those of most other insect orders [8, 21]. In addition, the components and divergence times of the various TE types are distinct among the different mantid lineages.

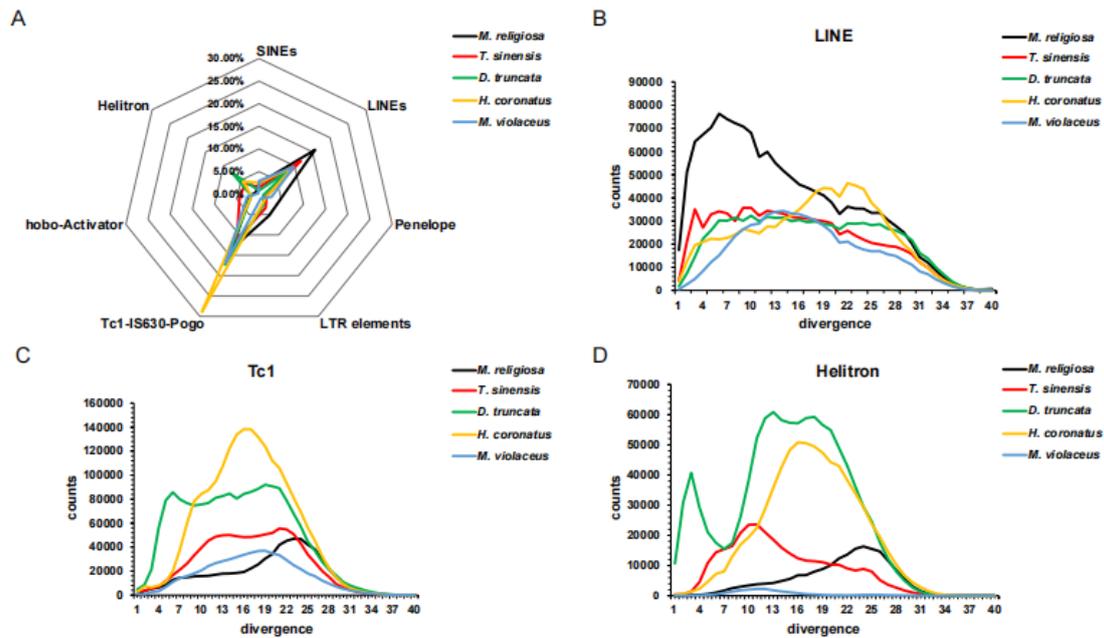

**Figure 2. TE distribution in five mantis genomes.** (A) The radar chart of component of TE. (B) The divergence distribution of LINE. (C) The divergence distribution of Tc1. (D) The divergence distribution of *Helitron*.

**Translocation between X chromosome and autosomes in the Mantoidea lineage**

Sex chromosomes evolved from autosomes and play important roles in tissue development, mating, and speciation [22-24]. The types of sex chromosomes found in insects vary among species, and sex chromosome systems exhibit significant diversity across insect species. Most insects have XX-XY, ZZ-ZW or XX-XO sex chromosome systems, but there are other rare sex chromosome types, such as the X1X2Y type, or two X chromosomes and one Y chromosome. Some hemipterans, including *Philaenus italosignus* [25] and mantids such as *Mantis religiosa* [26], exhibit this type.

To identify X chromosomes from the assembled pseudochromosomes, we generated 15X short-read sequencing data for female and male Chinese mantis individuals. Sequencing coverage revealed that all 14 chromosomes in females had comparable coverage depths, whereas in males, two chromosomes had approximately half the coverage depth (Figure 3A). It has been reported that most members of the superfamily Mantoidea have two X sex chromosomes, X1 and X2, derived from fusion or translocation between the X chromosome and an autosome [14]. The two chromosomes with half coverage depths are thus concluded to be the two sex chromosomes X1 and X2. The X1 and X2 chromosomes are the largest and second largest of our assembled pseudochromosomes, which is consistent with previous reports based on karyotyping [14, 27].

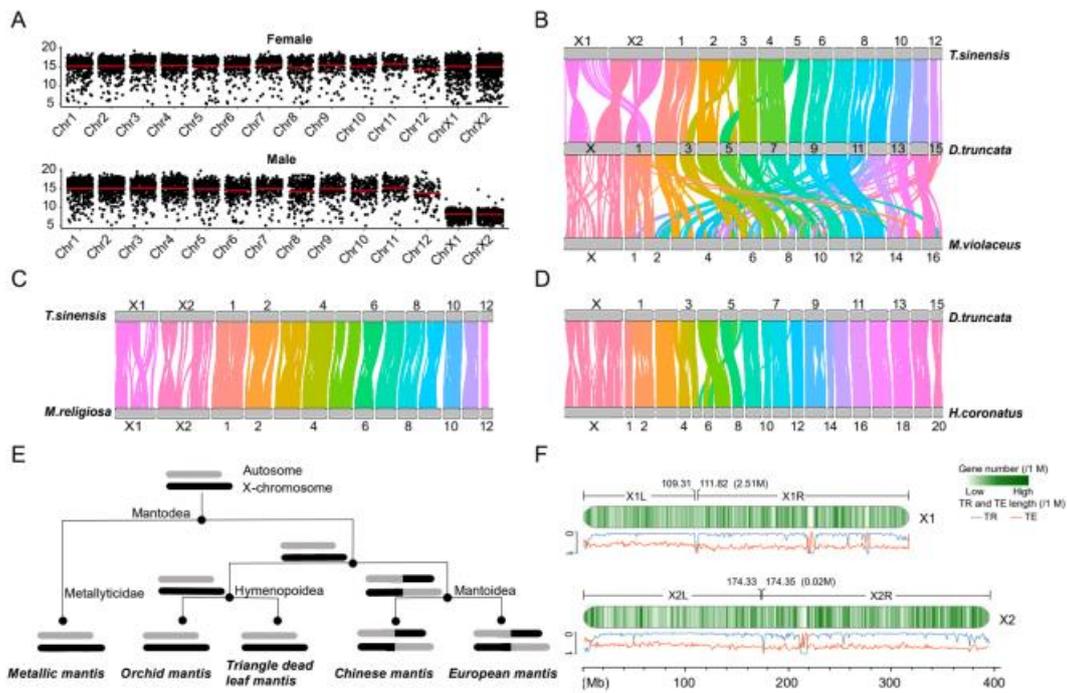

**Figure 3**. **Evolution of X chromosome in Mantodea.** (A) Identification of X chromosome in *T. sinensis* by comparing between male and female individual. The sequencing depth distributions were plotted in 500 Kb windows. The red line represents the average sequencing depth for each chromosome. (B) The synteny band plot among *T. sinensis*, *D. truncata* and *M. violaceus*. (C) The dual synteny between *M. religiosa* and *T. sinensis*. (D) The dual synteny band plot between *D. truncata* and *H. coronatus*. (E) The diagram shows the evolutionary process of the X chromosome along various lineages of Mantodea. (F) The range of the broken site for translocation on the X1 and X2 chromosomes of *T. sinensis*, which are 2.51 Mb and 0.02 Mb, respectively.

Macroscale synteny analysis was used to identify the corresponding X chromosomes in the other 4 species and allowed comparative analysis among the mantid X chromosomes. Synteny alignments revealed that both Chinese and European mantises have two sex chromosomes, X1 and X2; however, the other species have only one sex chromosome X. In addition, only parts of X1 (X1L) and X2 (X2L) in Mantoidea were aligned with the X chromosomes of the other 3 species (Figure 3B-D, S8). These results suggest that the ancestral mantid had one X chromosome and that the translocation of large fragments between the X chromosome and an autosome occurred in Mantidae (Figure 3E). Furthermore, based on conserved sequence alignments, we were able to identify the breakpoint range as a site falling within a 2.5-Mb region on the X chromosome (Figure 3F, S9). Resolution by sequencing is much greater than that previously obtained via cytological techniques, such as C-banding, silver staining and living-cell images of the meiosis process [14, 27].

Previous studies have revealed that the common ancestor of Dictyoptera had an XX-XY sex chromosome system [14, 27], in which females have two X chromosomes, whereas males have only one X chromosome (Figure S10). We infer that the common ancestor of the Mantidae family evolved two X chromosomes (X1 and X2), and we confirmed that the evolution of the X1 and X2 chromosomes resulted from the fusion and fragmentation of one X chromosome and an autosome. Our results will advance the evolution of sex determination systems and elucidate the mechanisms by which chromosome behaviour drives meiosis.

**The metallic mantis has evolved more slowly than the other mantises**

Comparative analysis of Mantodea genomes within a phylogenetic context is essential for understanding their evolution and diversity. Phylogenomic analyses were performed on these 5 Mantodea species, which span 5 genera and 3 families with diverse habitats and morphologies. Two Blattodea species, the German cockroach (*Blattella germanica*) [17] and the dampwood termite (*Zootermopsis nevadensis*) [21], were used as the outgroup (Table S8). From gene family clustering, 69,603 orthologous groups (OGs) were generated, including 4,014 single-copy OGs.

The metallic mantis belongs to the superfamily Metallyticoidea and shares many characteristics with its cockroach relatives but exhibits significant morphological differences compared to other mantises, including dull body colouration, a prostrate body posture, and a relatively shorter thorax. The Metallyticoidea lineage is sister to the other mantis lineages sequenced to date [28]. The metallic mantis shares more OGs with cockroaches than with the other 4 species do (Figure 4A), which may explain its strong morphological resemblance to cockroaches. Moreover, the phylogenetic tree was constructed based on 4,014 single-copy OGs, and the evolutionary rates along branches were estimated. The substitution rate for the metallic mantis branch was the lowest among those of Mantodea (Figure 4B), which may indicate that the evolution rate of the metallic mantis branch was slower than that of the other mantises. With a slower evolution rate, the metallic mantis may preserve more characteristics of the Mantodea ancestor, which further explains its morphological resemblance to cockroaches.

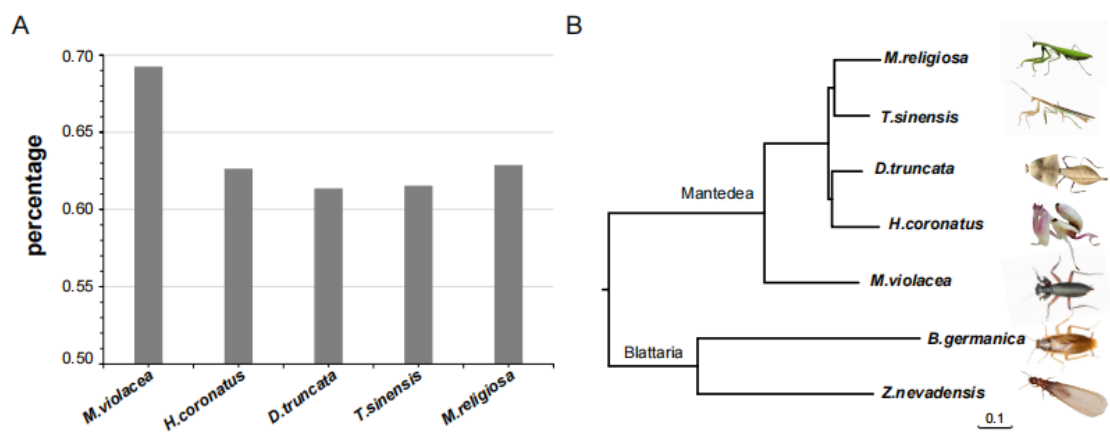

**Figure 4. Metallic mantis evolves slower than other mantises**. (A) Percentage of orthologous groups (OG) shared with cockroach for each mantis species. (B) Phylogeny is based on codon alignment of single copy genes (mantises, cockroaches, and termites). The branch length is in proportional with the substitution rate.

**A rapid radiation event in Mantodea followed the Cretaceous–Palaeogene (K–Pg) mass extinction event**

In the current NCBI Taxonomy database, the dead leaf mantis, European mantis, and Chinese mantis are classified into the same family, Mantidae (Figure 5). In this study, a phylogenetic tree based on single-copy orthologues in these 5 Mantodea species via both neighbour-joining and maximum likelihood methods revealed that the orchid mantis and triangle dead leaf mantis are close to each other (Figure S11), which differs from the findings of previous studies [1, 11, 29]. After adding the genomic data for *D. lobata*, both Deroplatys species sistered to the orchid mantis (Figure S12). In addition, we found more collinearity blocks between the orchid mantis and triangle dead leaf mantis (Figure 3D), implying more similar genomic structures.

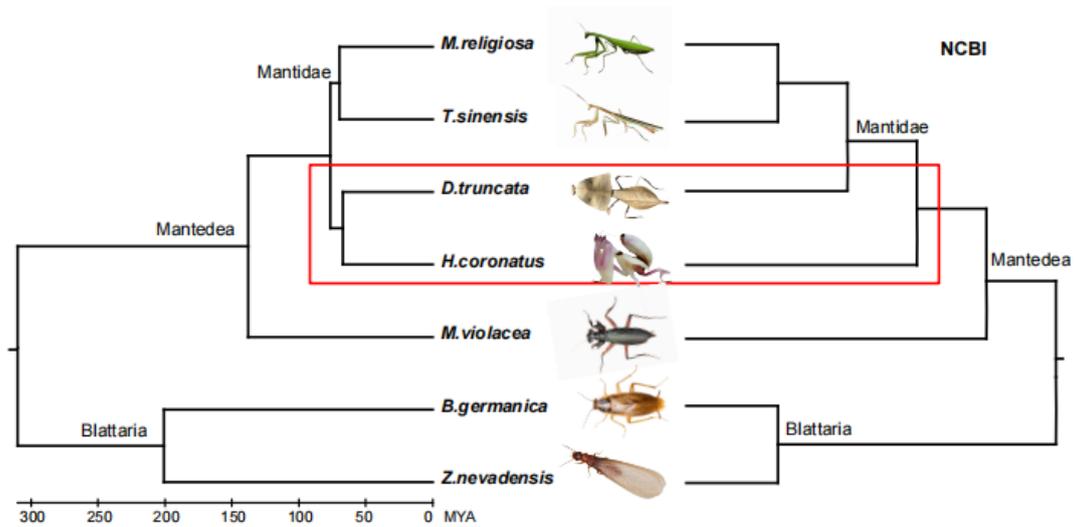

**Figure 5. The phylogeny of praying mantises.** Left part: phylogenetic relationships and divergence times based on 6,988 single copy genes (mantises); Right part: Phylogenetic relationships from NCBI. The position of *D. truncata* and *H. coronatus* was marked with a red frame. These trees have 100% bootstrap support for all evolutionary nodes.

Based on the phylogenetic tree, the divergence dating results revealed that the divergence of Mantodea and Blattodea occurred at ~320 Ma and that the entirety of mantis evolution spans more than 140 million years (Figure 5), which is mostly consistent with previous arthropod phylogenetic studies [30, 31]. Furthermore, we found that the divergence of the European mantis, Chinese mantis, triangle dead leaf mantis and orchid mantis, occurred within a short period of time at ~65 MYA (Figure 5), which closely followed the Cretaceous-Palaeogene (K-Pg) mass extinction event. A rapid radiation event may result in a controversial phylogeny [32], and the limited conservative genes and mitochondrial gene data further complicates the classification of mantis species. In this case, whole-genome data are necessary to obtain more convincing phylogenetic results and species classification.

**Discussion**

In this study, we generated chromosome-level genome assemblies for 5 mantis species via a combination of PacBio HiFi and Hi-C sequencing technologies. For the orchid mantis and Chinese mantis, both the contig N50 and N90 sizes of our assembly are approximately 5 times greater than those of the previously published reference genomes [12, 13]. According to our results, assembly continuity for European and Chinese mantises is relatively lower than that for the other 3 mantises due to the differences in heterozygosity, suggesting that tissues with no or low heterozygosity are still a priority for *de novo* genome sequencing. Compared with those of cockroaches and termites, the much larger genome sizes of mantises are mainly the result of expansions of various types of transposable elements.

Mantodea has occupied an important position in the evolution of insects, and one of its major sublineages, the superfamily Mantidae, has a special X1X2Y sex determination system. Through comparative genomics analysis, we inferred that the mantid common ancestor had only one X chromosome and that translocation between the X chromosome and an autosome occurred in the ancestor of Mantidae. We were able to narrow this breakpoint to less than a 2.5-Mb range on the original X chromosome, which will facilitate studies of sex determination systems and chromosome evolution. In the early diverging mantid lineage Metallyticoidea, the *Metallyticus* mantis genome shares more orthologous genes with cockroaches than with the other mantises, and the mutation rate of the metallic mantis genome is the lowest among all the mantises. Based on the phylogeny constructed from single-copy orthologous genes and whole-genome-wide syntenic patterns, we have some doubts about the accuracy of the previous classification of the triangle dead leaf mantis within Mantidae. The divergence of the European mantis, Chinese mantis, triangle dead leaf mantis and orchid mantis occurred within a short period of time at ~65 MYA, which closely followed the Cretaceous-Palaeogene (K-Pg) mass extinction event, indicating that the entirety of order Mantodea may have undergone rapid radiation after the K-Pg mass extinction event, which may also have contributed factor to the confusion in species classification.

Praying mantises have many other unique and interesting characteristics that are worth investigating, such as the formation of the cyclopean ear, the transformation of the foreleg, and camouflage. Although praying mantises are efficient predators, their hunting objects are not specific to harmful insects, hindering their wide application in organic planting. Thus, the genomic resources generated in this study will promote the biological analysis and further understanding of the evolution of Mantodean species.

## Methods

### Insect collection and sequencing

Mantis adults were collected at different locations: *M. religiosa* and *T. sinensis* from the forest of Guangzhou, China; *H. coronatus* from the rainforest of Xishuangbanna, China; and *D. truncata* and *M. violaceus* from two captive breeding centers in Beijing, China. All mantis samples for sequencing had the intestine removed to avoid contamination by bacteria, fungi, and residual prey bodies. All the tissues were cleaned with 30% ethanol and ddH$_2$O, and then immersed in liquid nitrogen for cryopreservation.

For Pacific Biosciences (PacBio) HiFi sequencing, libraries with ~15 kb insert sizes were constructed from a female adult of every mantis, and sequenced on a PacBio Sequel II system (RRID: SCR_017990). Subreads were generated with an N50 size of 14.5 kb, and consensus reads (CCS reads) were generated via ccs software (v.3.0.0) [33] with the following parameters: -min-passes 0 -min-rq 0.99 -min-length 100 -max-length 50,000. For Illumina sequencing of a male adult of *T. sinensis*, a short paired-end DNA library with a 400 bp insert size was constructed via standard Illumina protocols and sequenced on an Illumina NovaSeq 6000 platform (RRID:SCR_016387).

Total RNA from the head, eye, thorax, abdomen, forefoot, midfoot and midfoot of female adults was extracted with TRIzol reagent (Invitrogen) and used to construct cDNA libraries. Transcriptome sequencing data were generated via the Illumina NovaSeq 6000 system in PE150 mode.

**Genome assembly and quality assessment**

K-mer frequencies from HiFi reads of five mantises were calculated via Kmerfreq (https://github.com/fanagislab/kmerfreq), and then genome sizes were estimated via GCE (GCE, RRID:SCR_017332). The PacBio HiFi reads were assembled into contigs via Hifiasm (v0.14) (Hifiasm, RRID:SCR_021069)[34] with the following parameters: -l 1 -s 0.7. To filter duplicated contigs in the assembly, purge_dups (v1.2.3) (purge dups, RRID:SCR_021173) [33] was adopted with the following parameters: -2 -a 50. The completeness of the assembly was evaluated using BUSCO (v5.2.2) (BUSCO, RRID:SCR_015008) based on the OrthoDB (v10) (OrthoDB, RRID:SCR_011980) Insecta database [35].

For Hi-C scaffolding, two strategies were applied. For *M. religiosa*, whose contigs are more fragmental, Hi-C reads were mapped to contigs via the Arima mapping pipeline (https://github.com/ArimaGenomics/mapping_pipeline), and then, YaHS (v1.2a.1) (YaHs, RRID:SCR_0229650) [36] was used to assemble the contigs into pseudo chromosomes. For the other four mantises, whose contigs are much larger, Hi-C reads were mapped to contigs by Bowtie 2 (v 2.2.2.7) (Bowtie 2, RRID:SCR_016368) [37], then HiC-Pro (v2.11.0-beta) (HiC-Pro, RRID:SCR_017643) [38] was adopted to identify valid ligation pairs and generate Hi-C link matrices among different contigs, and finally, the contigs were clustered, ordered, and oriented into pseudo-chromosomes using EndHiC (v1.0) (EndHiC, RRID:SCR_022110) [39] based on the Hi-C linkage information among contig ends.

**Genome annotation**

A *de novo* transposable element (TE) library was constructed with RepeatModeler (v2.0.2) (RepeatModeler, RRID:SCR_015027) with the parameters -engine ncbi-database [40], and then RepeatMasker (v4.1.0) (RepeatMasker, RRID:SCR_012954) was used to identify TEs in the reference genome, using both the *de novo* TE library and the public Repbase TE library (v26.05) (Repbase, RRID:SCR_021169). The tandem repeat elements in the genome were subsequently identified using Tandem Repeats Finder (TRF) (Tandem Repeats Finder, RRID:SCR_022193) (v4.09) [41].

The protein-coding gene models were annotated in two rounds. In the first round, the genes were predicted by integrating evidence from *de novo* gene predictions and transcriptome-based gene predictions. *De novo* gene prediction was performed on the TE-masked genome assembly with AUGUSTUS (v3.4.0) (Augustus, RRID:SCR_008417) [42]. For transcriptome-based gene prediction, the RNA-seq data were filtered by Fastp (v0.23.1) (fastp, RRID:SCR_016962) [43]and then mapped to the genome using Bowtie2 (v2.2.7) [37], and StringTie (v1.3.3b) (StringTie, RRID:SCR_016323) was then used to construct the gene models [44]. All the gene models obtained via the above two approaches were subsequently integrated with EVidenceModeler (v1.1.1) (EVidenceModeler, RRID:SCR_014659) [45]. In the second round, for each mantis, the protein sequences from the other 4 mantises were mapped to this genome assembly with Exonerate (v2.4.0) (Exonerate, RRID:SCR_016088) [46], and incomplete gene models were filtered. Finally, for each mantis, the *de novo* gene predictions, the transcriptome-based gene predictions, and the homology-based gene predictions were integrated with EVidenceModeler (v1.1.1) to generate a high-confidence and nonredundant gene set.

The completeness of the gene sets was assessed using BUSCO based on OrthoDB (v10) for Insecta. For gene functional annotation, the mantis protein sequences were aligned to the KEGG (KEGG, RRID:SCR_012773), eggNOG (eggNOG, RRID:SCR_002456), NR, and UniProt (SwissProt) databases using DIAMOND (v0.9.24.125) (DIAMOND, RRID:SCR_009457) [47], and only the best hits with E-values less than $1e^{-5}$ were retained. Moreover, InterProScan (v5.38) (InterProScan, RRID:SCR_005829) was used to annotate the protein domains and GO (Gene Ontology) terms [48].

**X chromosome identification and analysis**

To identify the X chromosome, the clean Illumina paired reads from female and male samples were mapped to the genome of *T. sinensis* via BWA (v0.7.17-r1188) (BWA, RRID:SCR_010910) [49]. The bam files were filtered using SAMtools (v1.6) (SAMTOOLS, RRID:SCR_002105) [50] with the parameters '-q 60 -F 1804', and paired reads mapped onto different chromosomes were also filtered. To assess the sequencing depth of each chromosome, SAMtools depth (v1.6) was used to calculate the average base coverage. The two chromosomes in males whose sequencing depth was approximately half that of the other chromosomes, were identified as X-derived chromosomes. For consistency with the karyotype results for *T. sinensis* [26] and *M. religiosa* [14], the larger one was denoted X2, whereas the smaller one was denoted X1.

Pairwise collinearity analyses were conducted using the protein sequences of five mantis species as markers. DIAMOND (v0.9.24.125) with the parameters 'blastp -f 6' was used to align the protein sequences of each species pair, and the reciprocal best pairs were used as inputs for MCScanX (MCScanX, RRID:SCR_022067) to identify syntenic blocks [51]. The inter species syntenic genomic blocks were visualized via the R package Ideogram [52]. Based on the collinearity alignments of the five mantises, the X chromosomes of the other four mantises were also identified. In addition, the translocation sites on chromosomes X1 and X2 were inferred from the collinearity alignment.

**Evolutionary analysis**

Seven Dictyoptera species, including the five mantises sequenced in this study, as well as the German cockroach (*B. germanica*) [17] and dampwood termite (*Z. nevadensis*) [21], were used to infer orthologous genes via OrthoFinder (v2.5.4) (OrthoFinder, RRID:SCR_017118) with the default parameters [53]. The protein sequences of single-copy genes from each species were multiple aligned using MAFFT (v7.487) (MAFFT, RRID:SCR_011811) and then concatenated into one super protein sequence. Using the concatenated super protein sequence, RAxML (v8.2.12) (RAxML, RRID:SCR_006086) was subsequently employed to construct a maximum-likelihood phylogenetic tree with the PROTGAMMALGX model, and codon alignment of the super protein sequence were used to construct a maximum-likelihood phylogenetic tree with the GTRGAMMA model [54]. MEGA X was adopted to construct a neighbour-joining phylogenetic tree [55]. Node support was assessed via a bootstrap procedure based on 100 replicates. Divergence times among species were calculated via the RelTime branch method in

MCMCTree (PAML package, v. 4.7; PAML, RRID:SCR 014932)[56]. The calibration times were set according to the knowledge in TimeTree: 212-225 MYA between *M. religiosa* and *B. germanica*, and 173-210 MYA between *Z. nevadensis* and *B. germanica*.

## Abbreviations

BLAST: Basic Local Alignment Search Tool; bp: base pairs; BUSCO: Benchmarking Universal Single-Copy Orthologs; BWA: Burrows-Wheeler Aligner; CCS: circular consensus sequencing; Gb: gigabase pairs; GO: Gene Ontology; kb: kilobase pairs; KEGG: Kyoto Encyclopedia of Genes and Genomes; Ma: megaannus; Mb: megabase pairs; MYA: million years ago; NCBI: National Center for Biotechnology Information; NR: Non-Redundant; OG: orthologous groups; PacBio: Pacific Biosciences; PE: Paired end; RAxML: Randomized Axelerated Maximum Likelihood; TRF: Tandem Repeats Finder; TE: transposable element; TPM: transcripts per million; YaHS: yet another Hi-C scaffolding.

## Acknowledgments

This work was supported by Shenzhen Science and Technology Program (Grant No. KQTD20180411143628272) and Projects subsidized by Special Funds for Science Technology Innovation and Industrial Development of Shenzhen Dapeng New District (Grant No. PT202101-02); Fund of Key Laboratory of Shenzhen (ZDSYS20141118170111640) and The Agricultural Science and Technology Innovation Program.

## Data availability

The genomic and transcriptomic sequencing reads have been deposited in NCBI-SRA under the accession PRJNA987019, PRJNA989593, PRJNA989036, PRJNA988270, PRJNA989282 for *M. religiosa*, *T. sinensis*, *D. truncata*, *H. coronatus* and *M. violaceus*, respectively. The corresponding genome assemblies and annotations have been deposited at NCBI-Genome under the accessions JAUKNK000000000,

JAUKNM000000000, JAUKNL000000000, JAUKNX000000000, JAUJEO000000000, and are also available at Figshare (10.6084/m9.figshare.23995398, 10.6084/m9.figshare.23995410, 10.6084/m9.figshare.23995152, 10.6084/m9.figshare.23988987, 10.6084/m9.figshare.23995434).

**Author contributions**

H.L. and L.L. prepared the sequencing samples, performed data analysis, and wrote the raw manuscript. W.F. and G.W. supervised the project and revised the manuscript. The other authors provided helpful suggestions, and all authors read and approved the final version of this manuscript.

**Competing interests**

The authors declare no competing interest.

# Supporting Information

## Supplementary figures

**Mantis religiosa**

*Tenodera sinensis*

*Deroplatys truncata*

*Hymenopus coronatus*

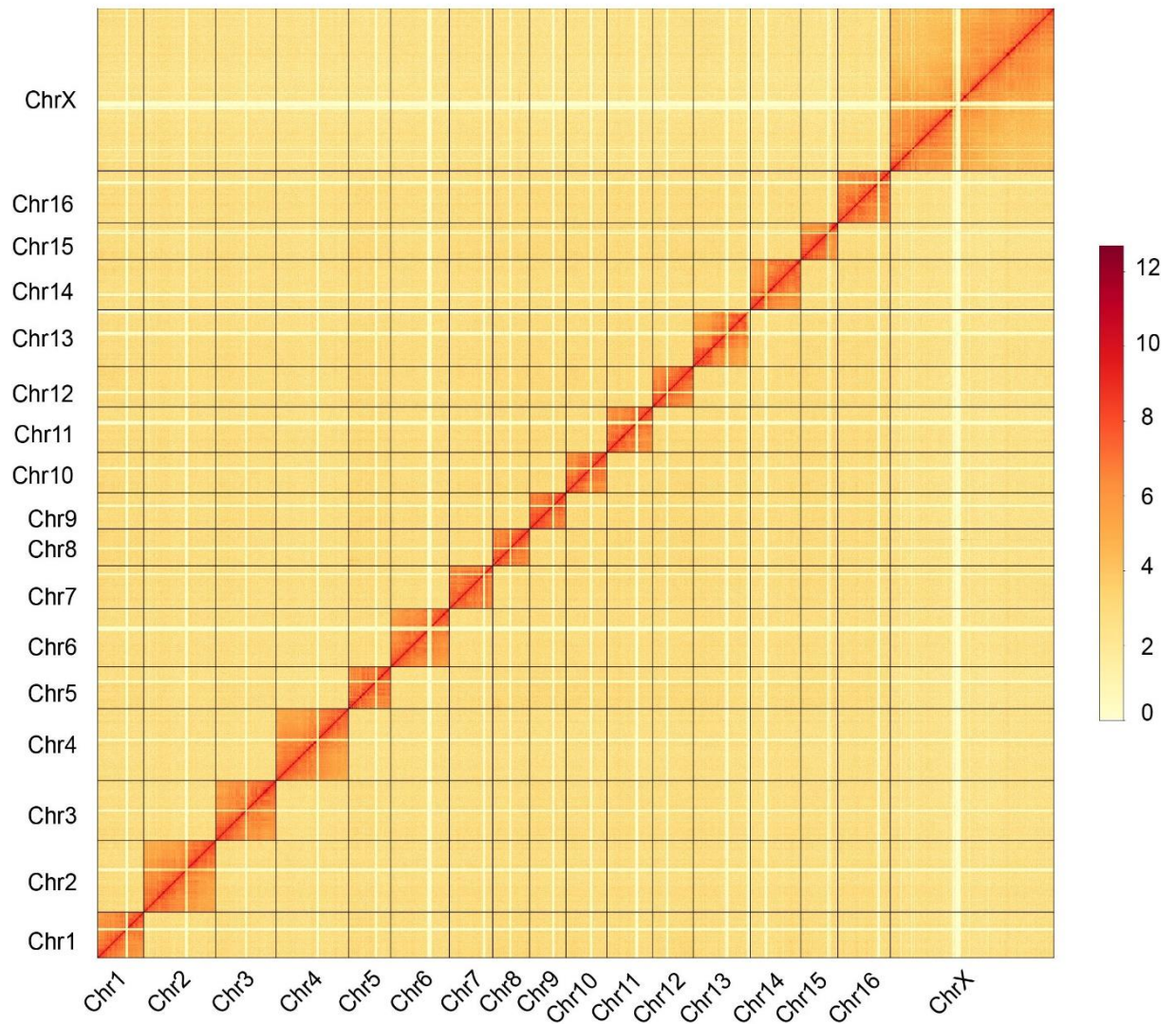

**Figure S1. Hi-C heatmap of genome assembly for *M. religiosa*, *T. sinensis*, *D. truncata*, *H. coronatus*, and *M. violaceus*.** The resolution (window size) is 1000-Kb, and color represents $\log_2$(Links number). Links number is the number of Hi-C links falling between the two analyzed genomic windows.

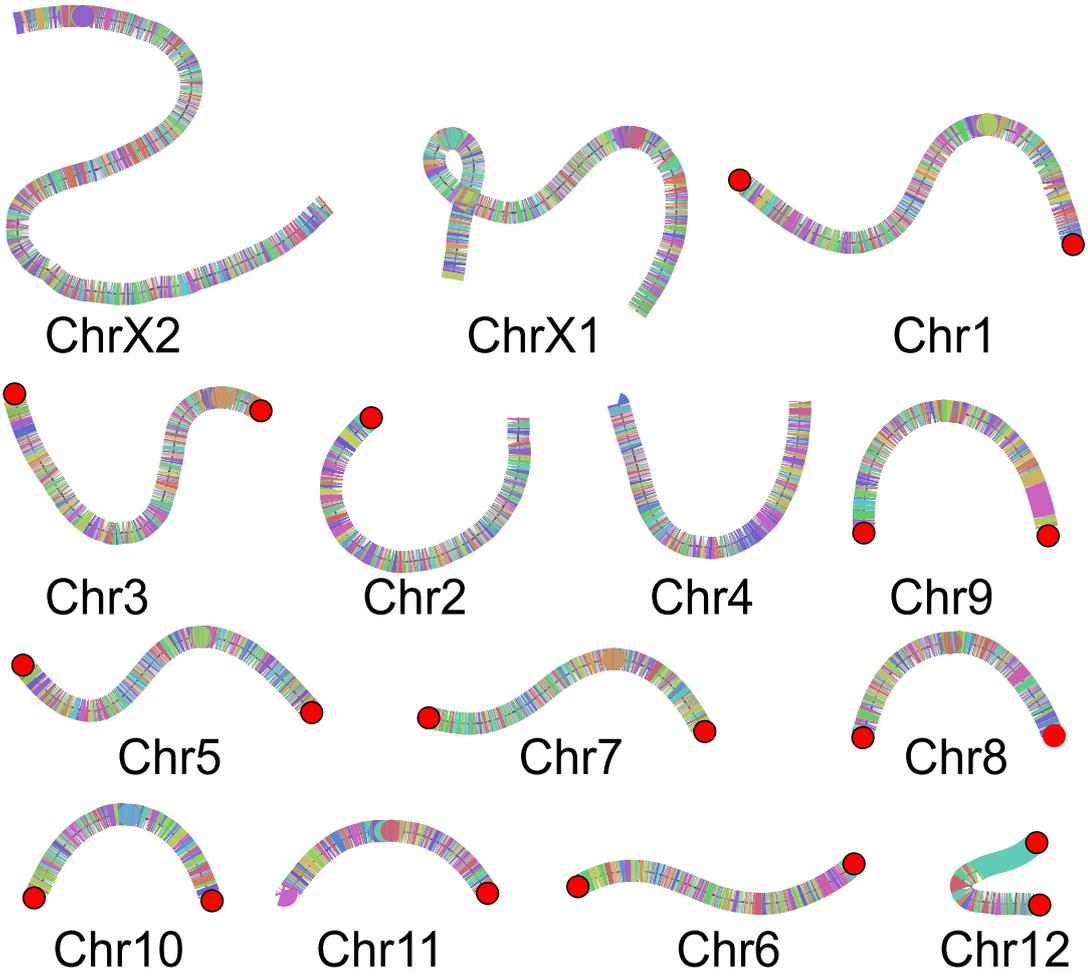

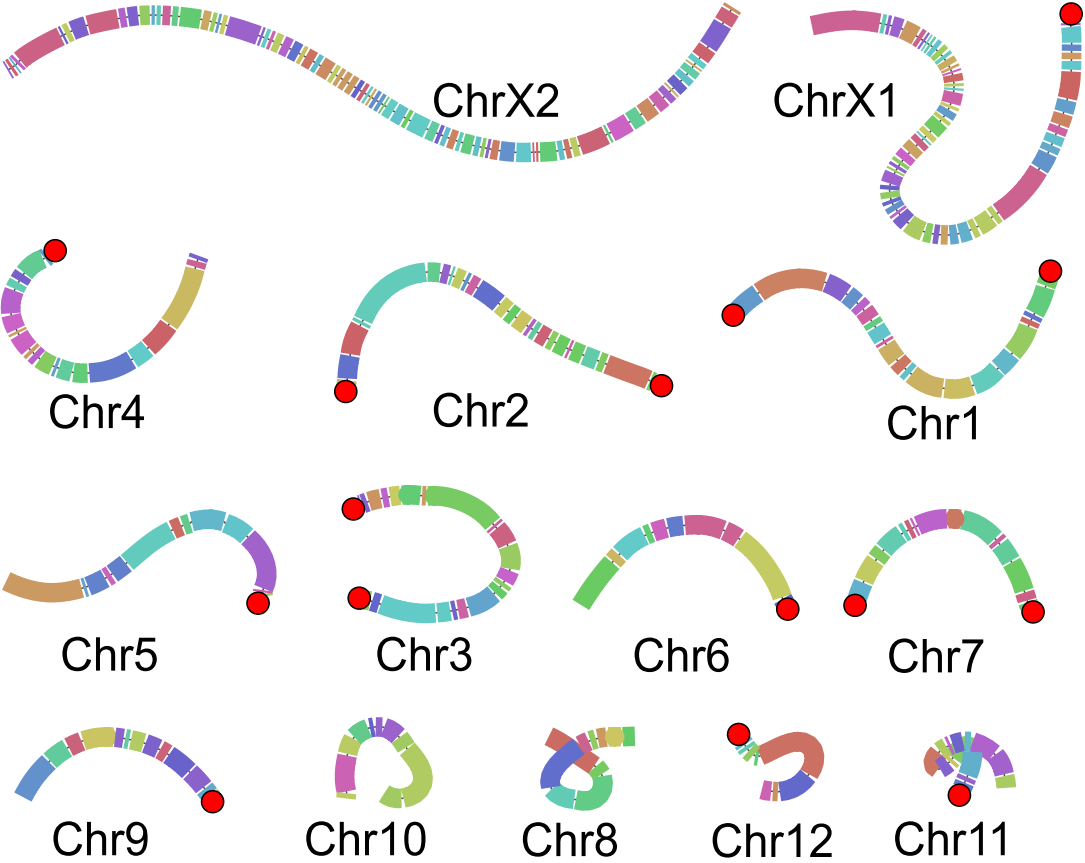

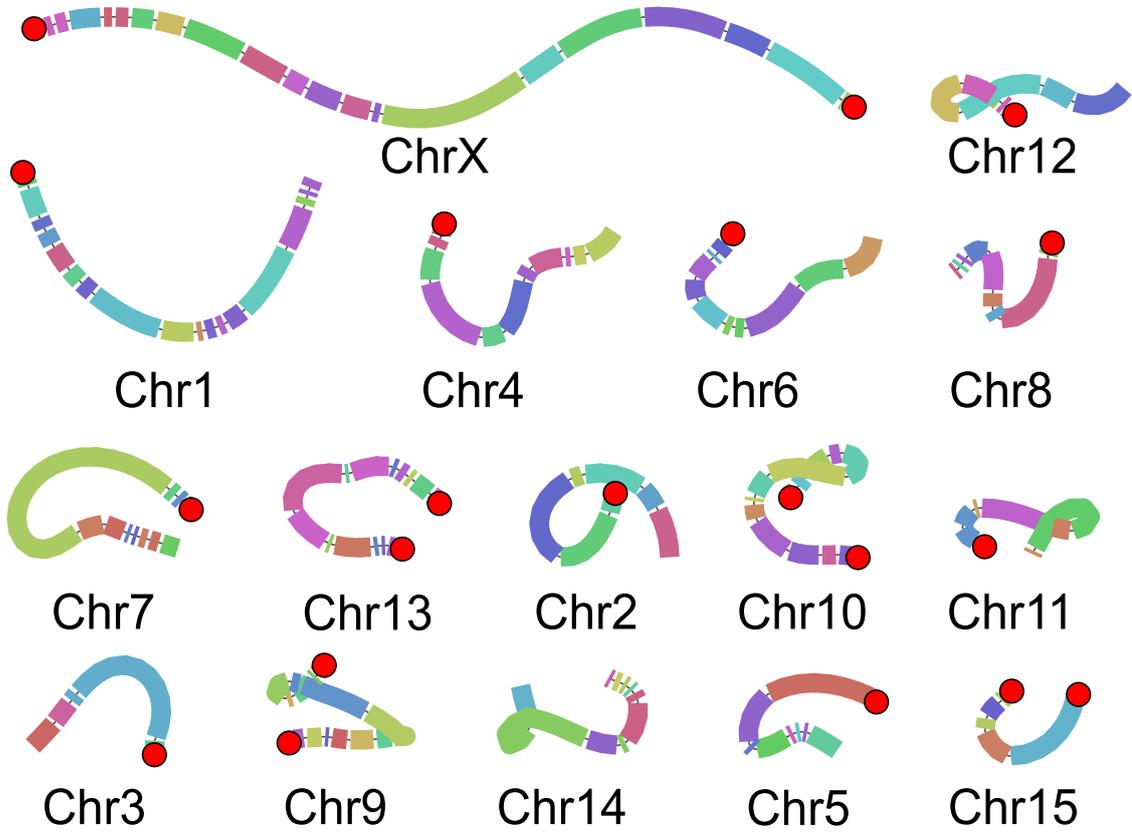

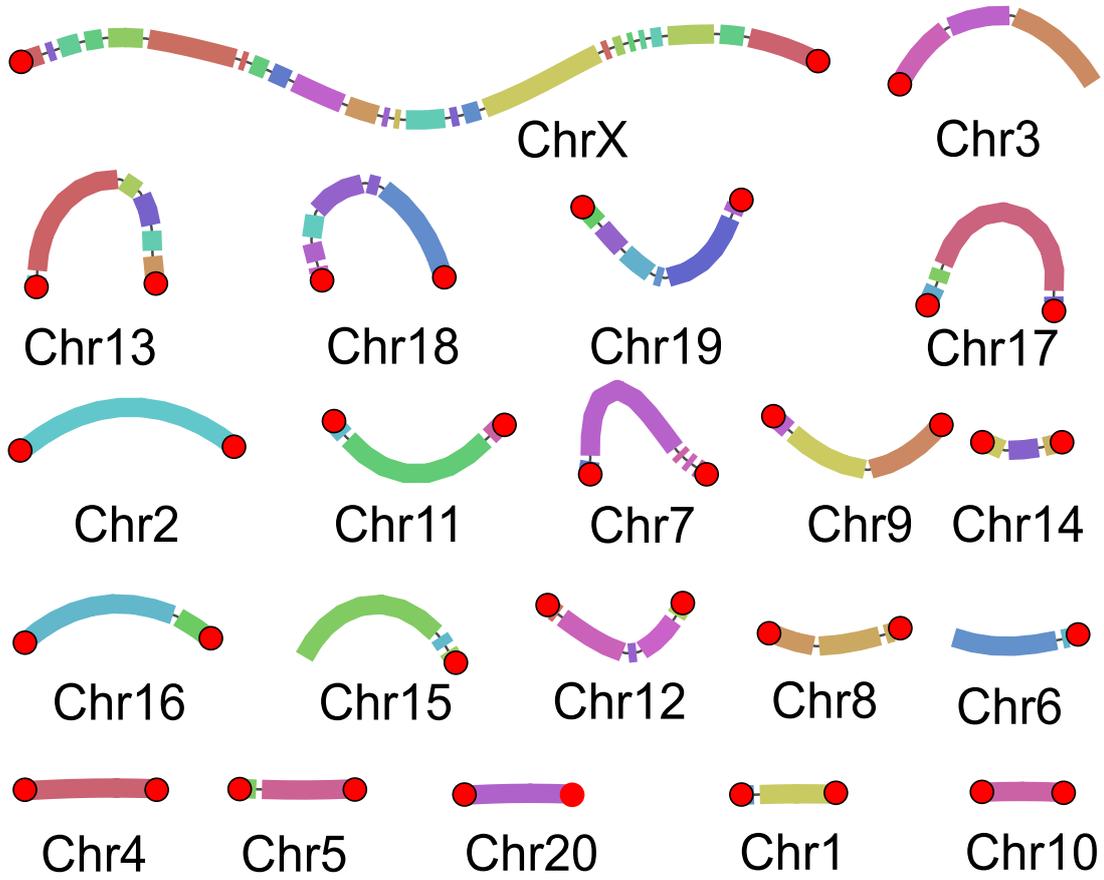

# *Metallyticus violaceus*

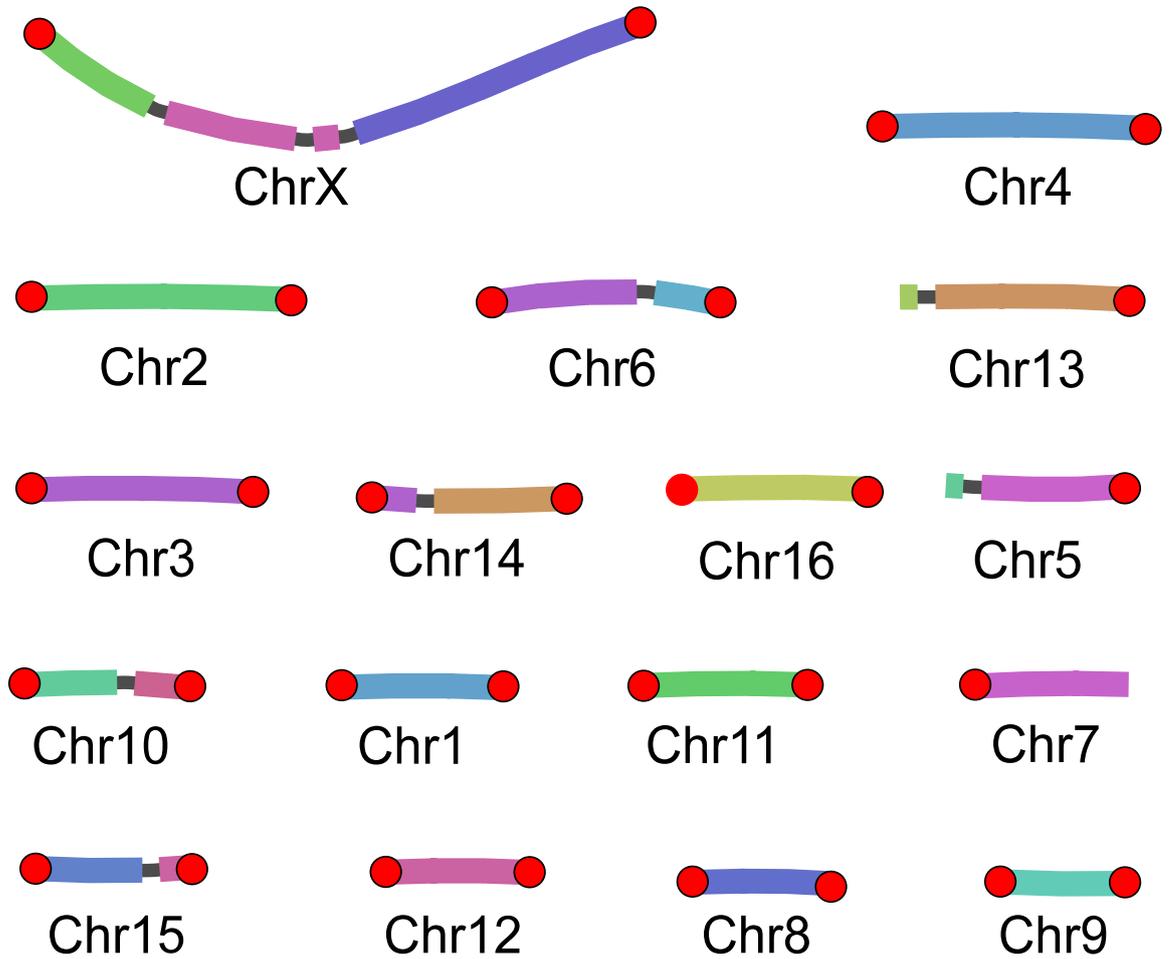

**Figure S2. Bandage view of scaffolding results for *M. religiosa*, *T. sinensis*, *D. truncata*, *H. coronatus*, and *M. violaceus*.** Each rectangle represents for a contig. The chromosome ends assembled with telomere-specific tandem repeats (unit: TTAGG) were highlighted with red circle.

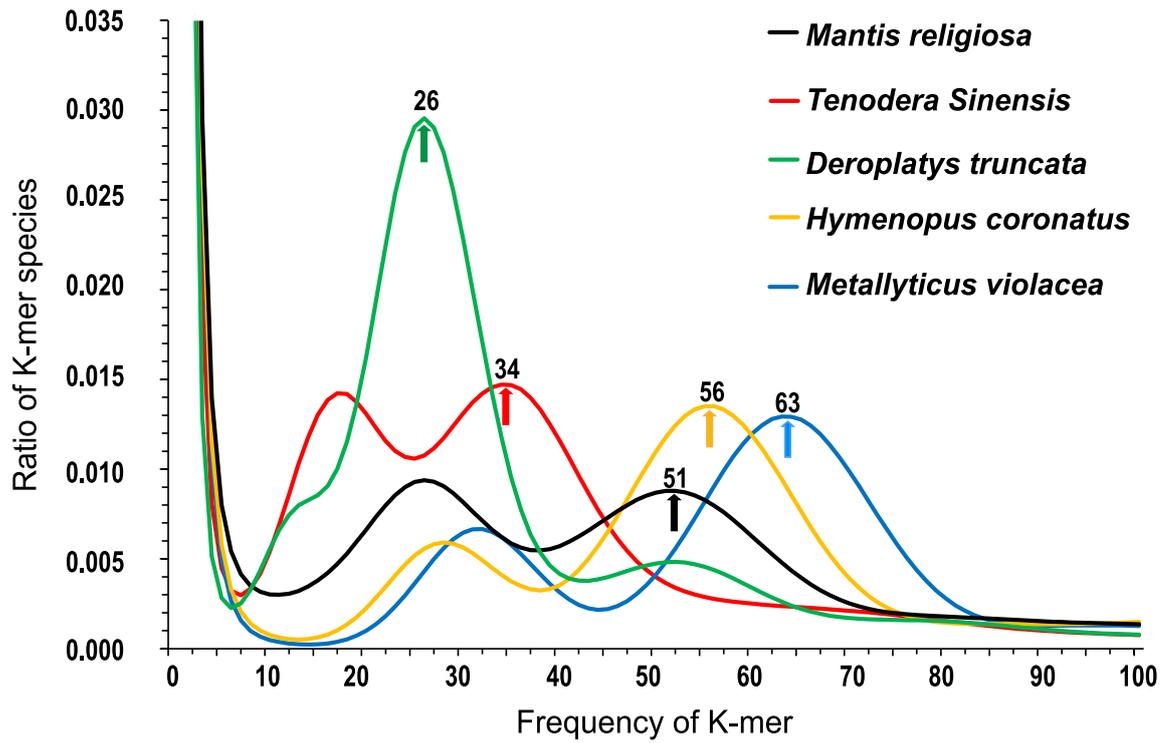

**Figure S3. Distribution of K-mer (K-size 17) frequencies in sequencing data for *M. religiosa*, *T. sinensis*, *D. truncata*, *H. coronatus*, and *M. violaceus*.** For each mantis, the left peak reflects the heterozygous regions, while the right peak reflects the homozygous regions, which were marked by arrow.

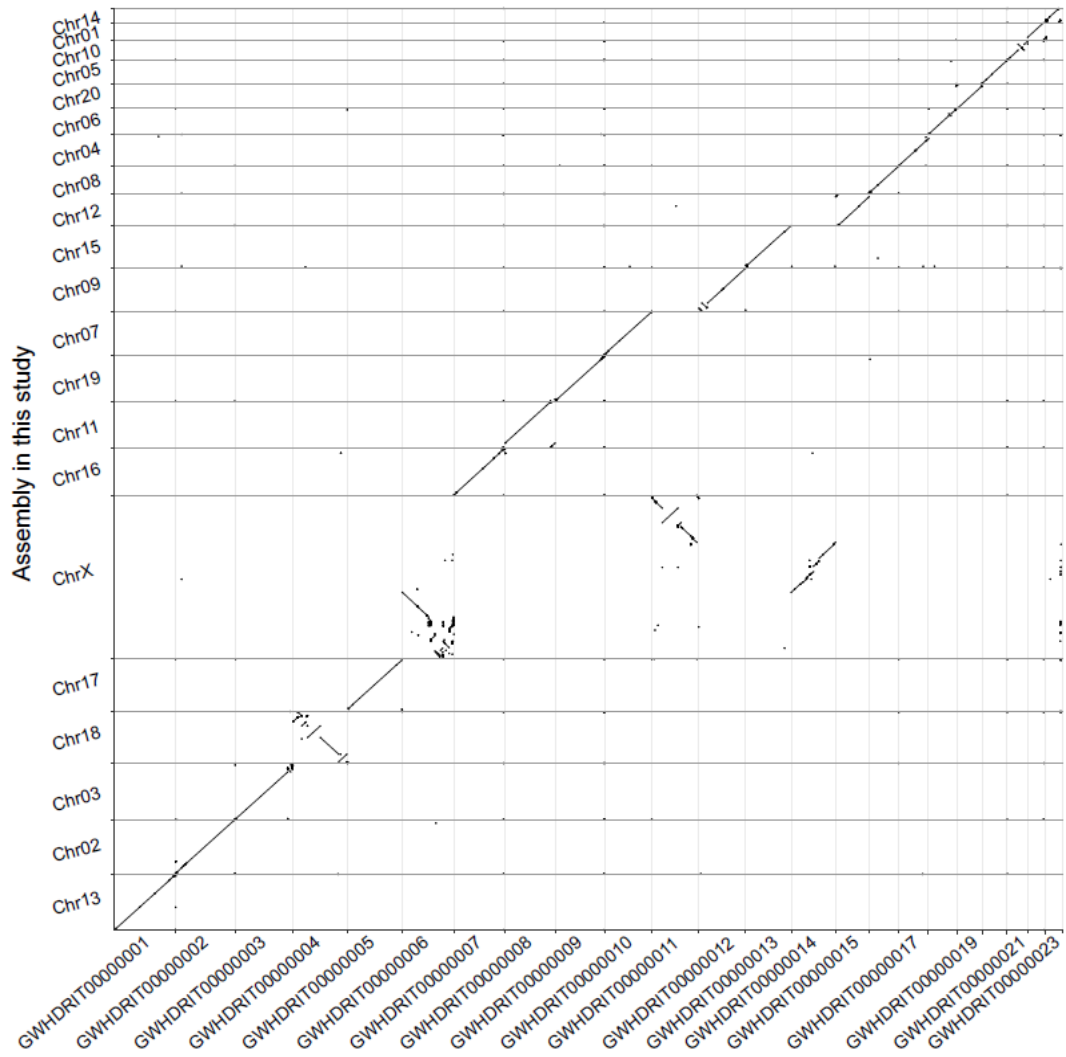

**Fig. S4. Syntenic comparison of our *H. coronatus* assembly with the published assembly from Huang, G *et al.*, 2023.** The whole genome alignment is performed by minimap2 with "-x asm10" parameter.

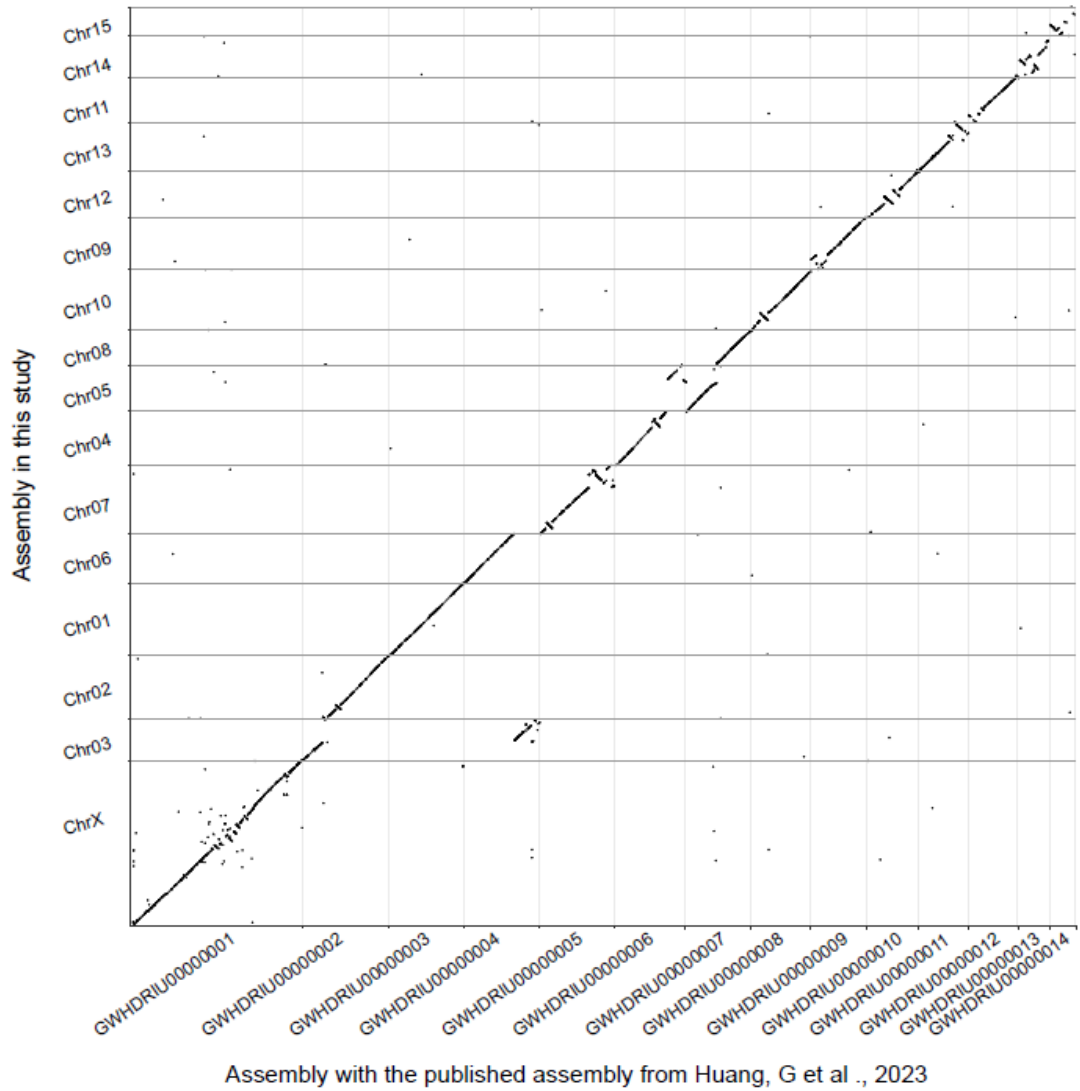

**Fig. S5. Syntenic comparison of our *D. truncata* assembly with the published assembly of *D. lobata* from Huang, G *et al.*, 2023.** The whole genome alignment is performed by minimap2 with "-x asm10" parameter.

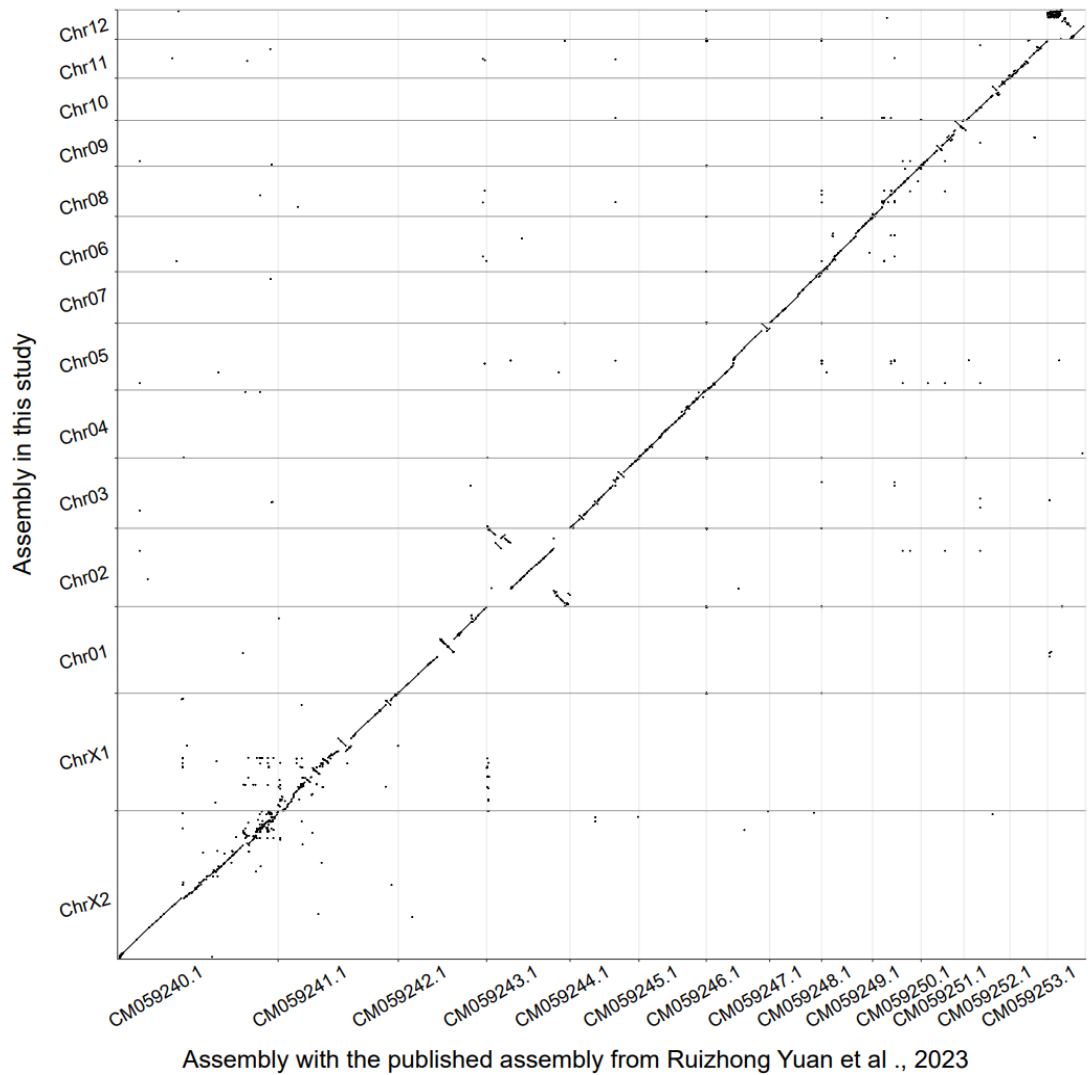

**Fig. S6. Syntenic comparison of our *T. sinensis* assembly with the published assembly from Ruizhong Yuan *et al.*, 2023.** The whole genome alignment is performed by minimap2 with "-x asm10"parameter.

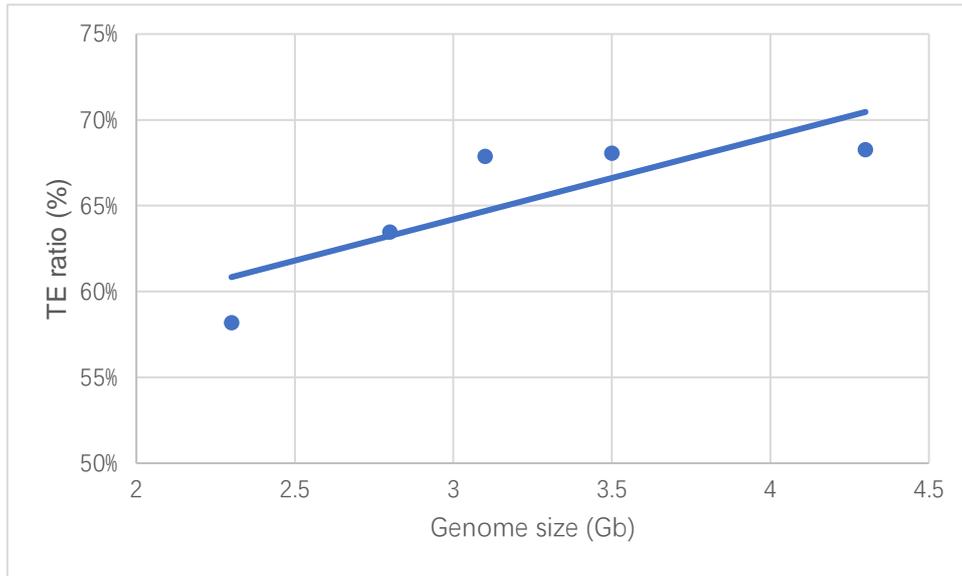

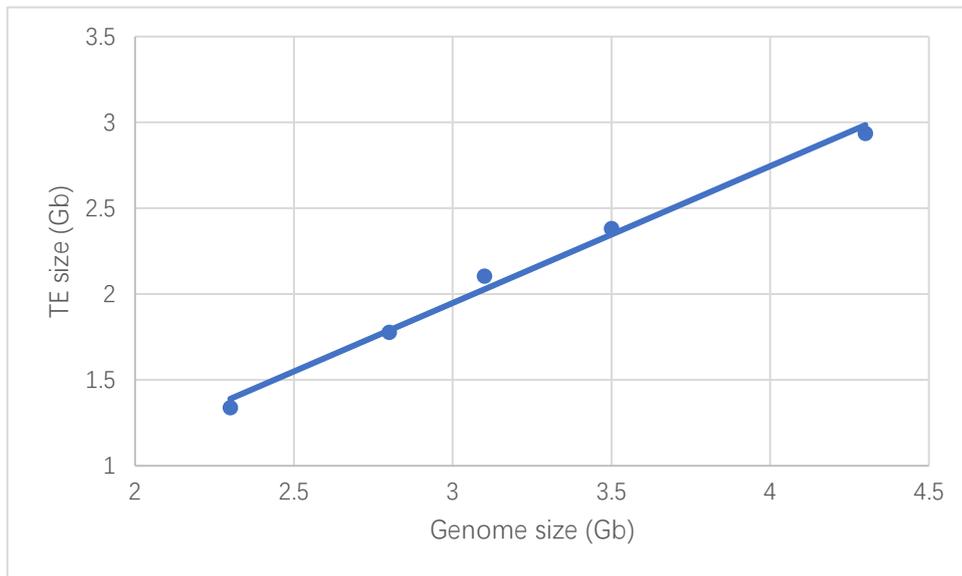

**Figure S7.** The relationship between genome size and TE size (ratio) of 5 Mantodea species.

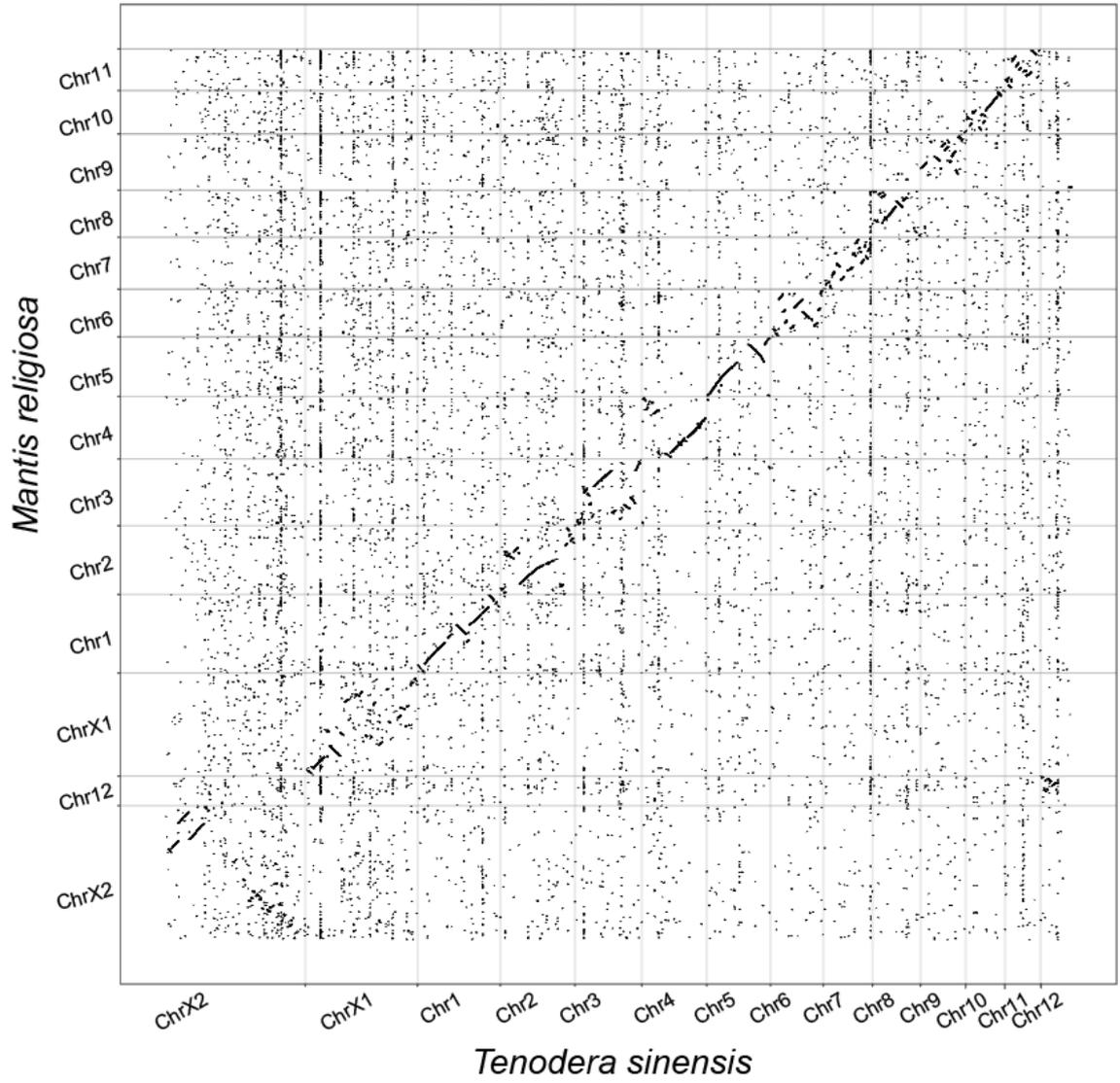

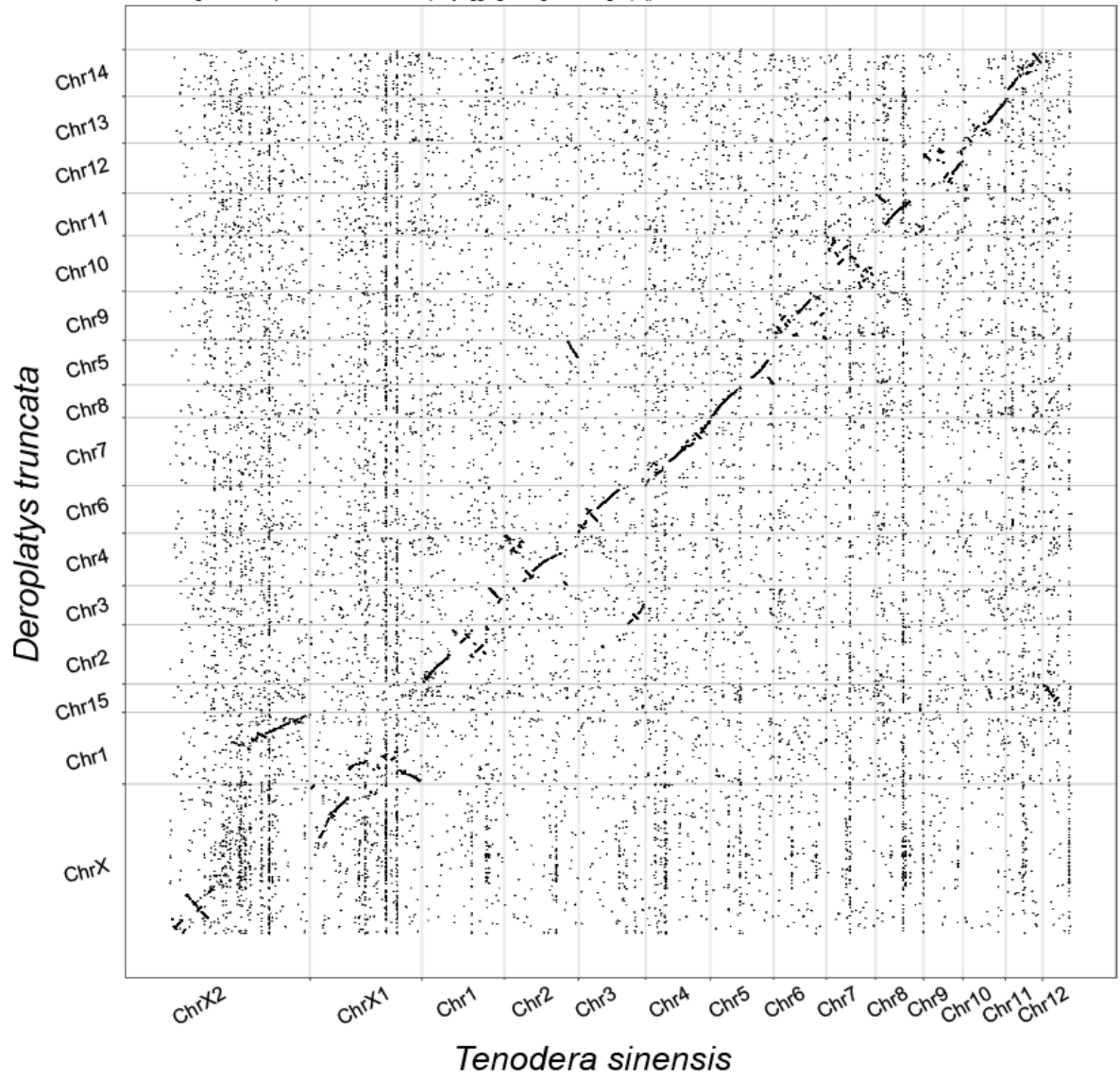

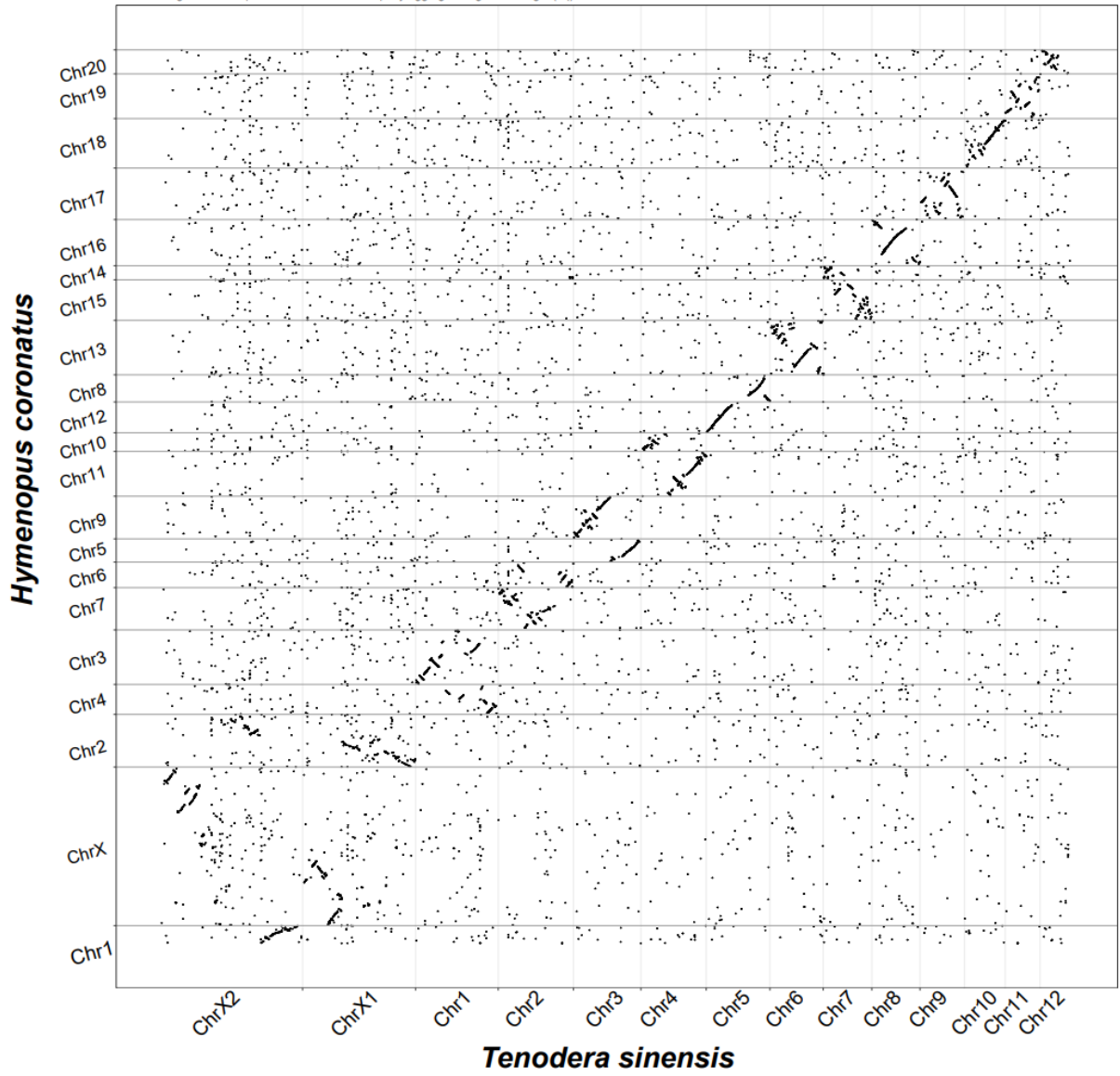

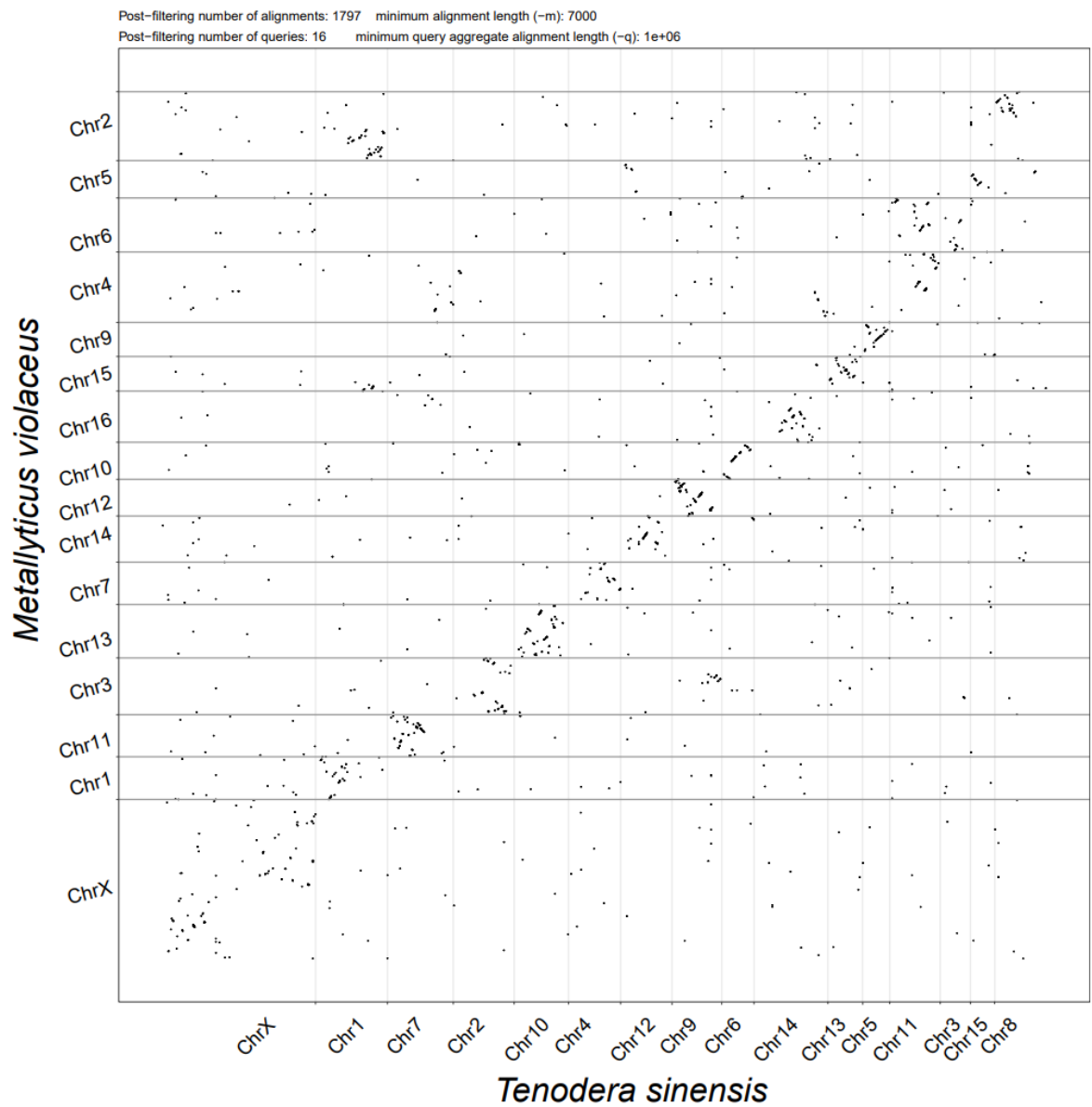

**Fig. S8. Syntenic comparison of *T. sinensis* with the other 4 mantises.** The whole genome alignment is performed by minimap2 with "-x asm10" parameter.

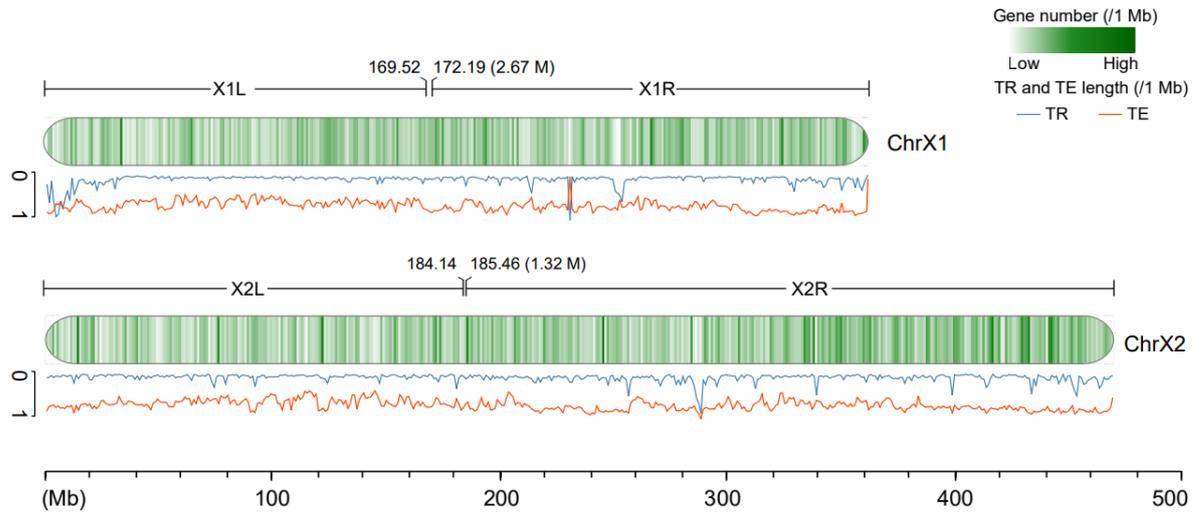

**Figure S9. The interval range for the translocation site on the X1 and X2 chromosomes of *M. religiosa*.** Distributions of TF (tandem repeat) and TE (transposable element) were shown.

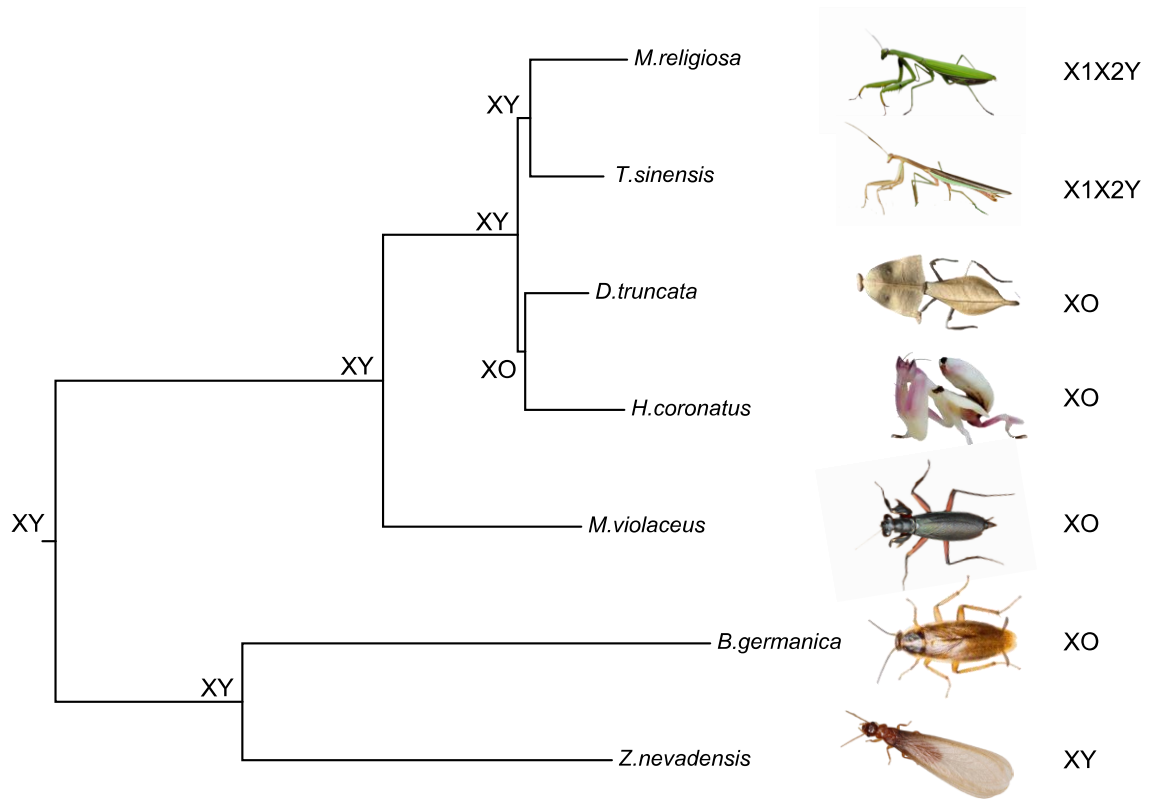

**Figure S10. Sex determination system of Mantodea (mantises) and Blattodea (cockroaches and termites).** The plot was draw refer to "evobir.shinyapps.io/PolyneopteraDB/".

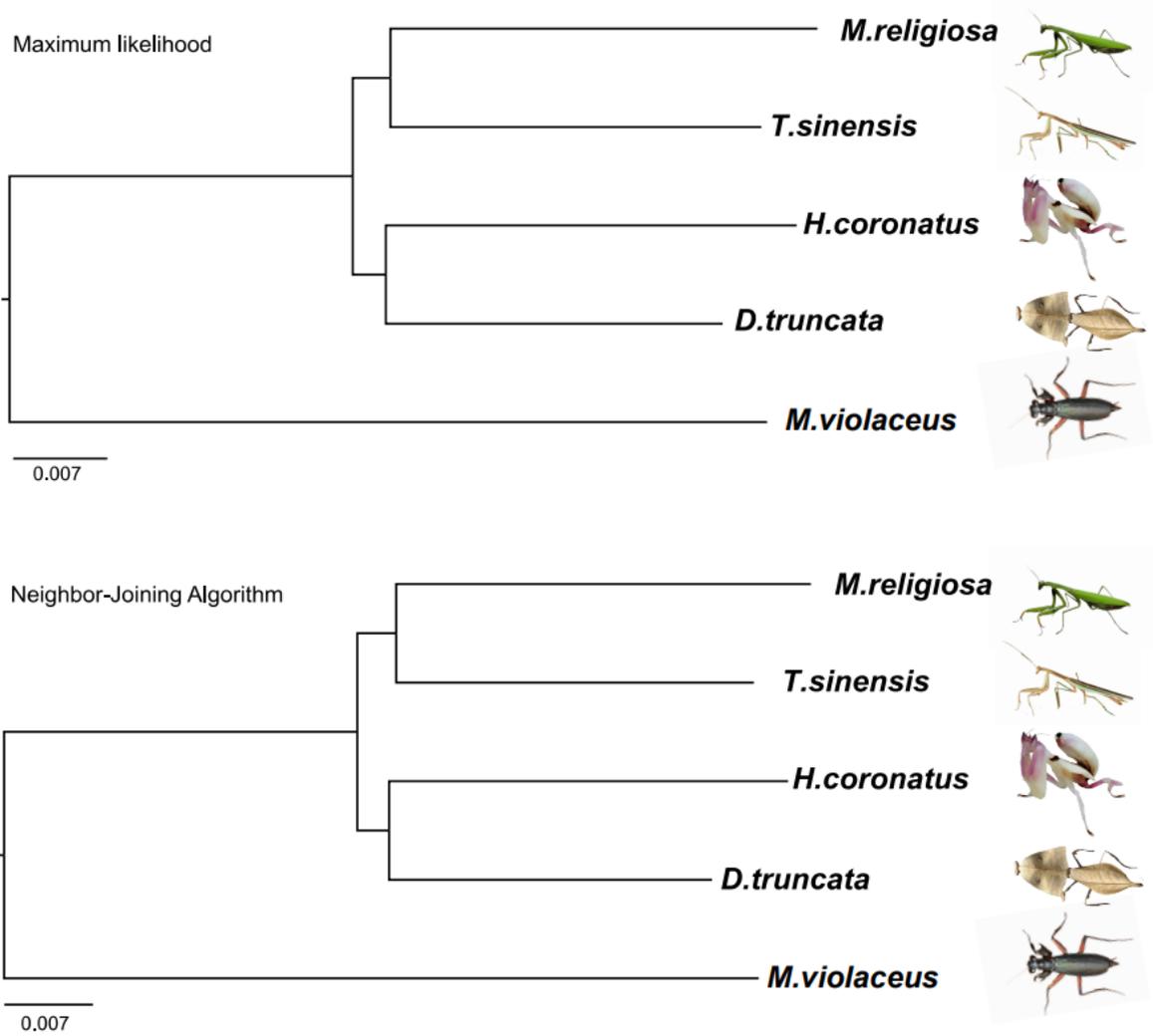

**Figure S11. Phylogeny tree for 5 mantis species.** The phylogenetic tree was built based on 6,988 single-copy orthologs, with both neighbor joining and maximum likelihood methods. The two phylogeny results are largely consistent.

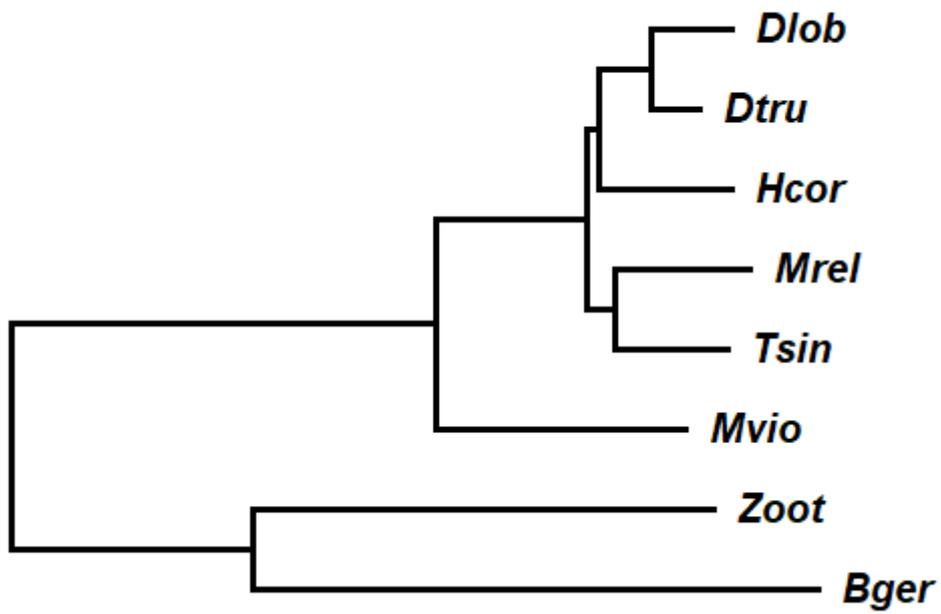

**Figure S12. Phylogeny tree for 8 Dictyoptera species.** The phylogenetic tree was built based on 2201 single-copy orthologs, with maximum likelihood methods. Dlob: *D.lobata,* Dtru: *D. truncate*, Hcor: *H. coronatus,* Mrel: *M. religiosa,* Tsin: *T. sinensis,* Mvio: *M. violaceus ,*Zoot: *Z. nevadensis,* Bger*: B. germanica*

# Supplementary tables

**Table S1.** Background information of the sequenced mantis species.

| Species name | Family / Sub-family | Chromosome number | source |
|---|---|---|---|
| *Mantis religiosa* | Mantidae / Mantinae | 2n = 28 | China: Guangzhou |
| *Tenodera sinensis* | Mantidae / Mantinae | 2n = 28 | China: Guangzhou |
| *Deroplatys truncata* | Mantidae / Deroplatyinae | 2n = 32 | Malaysia: Cameron |
| *Hymenopus coronatus* | Hymenopoidea / Hymenopodidae | 2n = 42 | China: Xishuangbanna |
| *Metallyticus violaceus* | Metallyticidae / Metallyticidae | 2n = 34 | Malaysia: Kuala Lumpur |

Note: The chromosome number of *D. truncata*, *H. coronatus*, and *M. violaceus* were determined by genome assembly. No karyotype information was available for these 3 mantis species.

**Table S2.** Statistics of genome sequencing data.

|  | Genomic Hifi (CCS) | | Genomic HiC | | Illumina RNAseq PE150 | |
|---|---|---|---|---|---|---|
|  | reads | bases | reads | bases | reads | bases |
| *Mantis religiosa* | 10,736,312 | 179,683,809,522 | 847,343,050 | 127,101,457,500 | 618,200,522 | 92,730,078,300 |
| *Tenodera sinensis* | 4,901,197 | 97,511,748,151 | 749,492,290 | 112,423,843,500 | 487,921,752 | 73,188,262,800 |
| *Deroplatys truncata* | 7,744,738 | 112,906,676,547 | 1,023,722,878 | 153,558,431,700 | 624,484,224 | 93,672,633,600 |
| *Hymenopus coronatus* | 10,167,730 | 177,528,078,708 | 1,214,765,322 | 182,214,798,300 | 630,950,136 | 94,642,520,400 |
| *Metallyticus violaceus* | 9,250,962 | 147,014,941,985 | 1,225,352,740 | 183,802,911,000 | - | - |

**Table S3.** Statistics of Hi-C data mapping to contigs.

| HiC pro results | *Mantis religiosa* | | *Tenodera sinensis* | | *Deroplatys truncata* | | *Hymenopus coronatus* | | *Metallyticus violaceus* | |
|---|---|---|---|---|---|---|---|---|---|---|
| | Reads number | Reads % | Reads number | Reads % | Reads number | Reads % | Reads number | Reads % | Reads number | Reads % |
| Total pairs processed | 423,671,525 | 100.0% | 374,746,145 | 100.0% | 511,861,439 | 100.0% | 607,382,661 | 100.0% | 612,676,370 | 100.0% |
| Unmapped pairs | 20,914,316 | 4.9% | 31,898,335 | 8.5% | 5,258,539 | 1.0% | 8,315,587 | 1.4% | 26,580,086 | 4.3% |
| Low qual pairs | 157,153,623 | 37.1% | 148,457,384 | 39.6% | 201,124,421 | 39.3% | 213,630,868 | 35.2% | 141,182,216 | 23.0% |
| Pairs with singleton | 80,162,675 | 18.9% | 123,773,101 | 33.0% | 44,326,963 | 8.7% | 73,893,690 | 12.2% | 85,201,198 | 13.9% |
| Unique paired alignments | 165,440,911 | 39.0% | 70,617,325 | 18.8% | 261,151,516 | 51.0% | 311,542,516 | 51.3% | 359,712,870 | 58.7% |
| Valid interaction pairs | 85,602,445 | 20.2% | 56,106,670 | 15.0% | 228,045,907 | 44.6% | 273,688,263 | 45.1% | 129,362,070 | 21.1% |
| Dangling end pairs | 73,543,590 | 17.4% | 9,246,952 | 2.5% | 28,174,610 | 5.5% | 27,962,589 | 4.6% | 205,423,801 | 33.5% |
| Religation pairs | 3,169,697 | 0.7% | 860,676 | 0.2% | 1,493,478 | 0.3% | 4,059,892 | 0.7% | 16,514,464 | 2.7% |
| Self Cycle pairs | 141,033 | 0.0% | 60,031 | 0.0% | 418,528 | 0.1% | 455,795 | 0.1% | 3,822,912 | 0.6% |
| Filtered pairs | 2,978,737 | 0.7% | 4,328,894 | 1.2% | 3,016,583 | 0.6% | 5,365,973 | 0.9% | 4,562,422 | 0.7% |
| Dumped pairs | 5,409 | 0.0% | 14,102 | 0.0% | 2,410 | 0.0% | 10,004 | 0.0% | 27,201 | 0.0% |
| valid interaction | 85,602,445 | 20.2% | 56,106,670 | 15.0% | 228,045,907 | 44.6% | 273,688,263 | 45.1% | 129,362,070 | 21.1% |
| valid interaction rmdup | 71,796,880 | 16.9% | 43,658,431 | 11.7% | 181,375,815 | 35.4% | 202,919,881 | 33.4% | 34,648,667 | 5.7% |

Note: The statistics numbers were obtained from Hi-C pro result files: *.mpairstat, *.mRSstat, and * allValidPairs.mergestat. The "valid interaction rmdup" represents non-redundant and valid Hi-C read pairs, which were used by EndHiC for scaffolding.

**Table S4.** Statistics of genome assembly.

| Species | Total contig size (bp) | Contig N50 size (bp) | Contig N90 size (bp) | Total scaffold size (bp) | Scaffold N50 size (bp) | Scaffold N90 size (bp) | % anchored to chromosomes | BUSCO results with insecta odb10 |
|---|---|---|---|---|---|---|---|---|
| *Mantis religiosa* | 3,675,712,721 | 1,407,320 | 339,785 | 3,680,002,721 | 210,326,877 | 3,517,080 | 85.39% | C:98.7% [S:94.1%, D:4.6%] |
| *Tenodera sinensis* | 2,687,116,722 | 13,794,865 | 2,923,070 | 2,687,426,722 | 190,002,057 | 104,358,734 | 95.63% | C:99.3% [S:95.6%, D:3.7%] |
| *Deroplatys truncata* | 4,290,634,545 | 44,541,363 | 9,432,610 | 4,290,792,545 | 248,405,437 | 184,535,799 | 97.47% | C:98.6% [S:96.6%, D:2.0%] |
| *Hymenopus coronatus* | 3,127,524,514 | 71,519,735 | 15,255,482 | 3,127,590,514 | 159,059,693 | 82,661,183 | 98.27% | C:99.0% [S:97.7%, D:1.3%] |
| *Metallyticus violaceus* | 2,322,120,794 | 109,157,195 | 61,664,404 | 2,322,129,794 | 125,733,329 | 88,780,966 | 98.51% | C:98.9% [S:97.4%, D:1.5%] |

Note: % anchored to chromosomes represents for percent of contig sequences assembled into chromosome-level scaffolds. In the BUSCO results, C means complete gene, S means single copy gene, while D means duplicated gene. S and D are both complete genes.

**Table S5.** Statistics of tandem repeats annotation.

| Species | Total scaffold size (bp) | Total TR size (bp) | TR rate % | TR N50 size (bp) |
|---|---|---|---|---|
| *M. religiosa* | 3,680,002,721 | 396,842,330 | 10.8% | 3,188 |
| *T. sinensis* | 2,687,426,722 | 403,304,947 | 15.0% | 30,747 |
| *D. truncata* | 4,290,792,545 | 471,243,565 | 11.0% | 35,332 |
| *H. coronatus* | 3,127,590,514 | 238,530,960 | 7.6% | 6,018 |
| *M. violaceus* | 2,322,129,794 | 186,949,249 | 8.1% | 2,372,101 |

Note: TR represents for "tandem repeat".

**Table S6.** Statistics of TE content in various classes.

| TE class | *Mantis religiosa* | | *Tenodera sinensis* | | *Deroplatys truncata* | | *Hymenopus coronatus* | | *Metallyticus violaceus* | |
|---|---|---|---|---|---|---|---|---|---|---|
| | Length (bp) | Percent | Length (bp) | Percent | Length (bp) | Percent | Length (bp) | Percent | Length (bp) | Percent |
| SINEs | 88,675,710 | 2.41% | 41,364,524 | 1.51% | 34,634,604 | 0.81% | 79,926,835 | 2.56% | 68,180,146 | 2.94% |
| Penelope | 179,087,096 | 4.87% | 46,171,123 | 1.69% | 40,398,887 | 0.94% | 56,917,120 | 1.82% | 65,477,245 | 2.82% |
| LINEs | 579,512,280 | 15.77% | 321,787,695 | 11.75% | 369,853,954 | 8.62% | 299,105,729 | 9.56% | 226,842,133 | 9.77% |
| LTR elements | 191,759,399 | 5.22% | 92,024,175 | 3.36% | 106,555,013 | 2.48% | 64,704,975 | 2.07% | 22,924,553 | 0.99% |
| DNA transposons | 775,059,780 | 21.09% | 521,921,270 | 19.06% | 1,039,790,255 | 24.23% | 1,023,083,723 | 32.71% | 516,358,740 | 22.24% |
| hobo-Activator | 80,830,494 | 2.20% | 120,355,965 | 4.39% | 72,581,984 | 1.69% | 53,762,543 | 1.72% | 62,477,753 | 2.69% |
| Tc1-IS630-Pogo | 487,487,116 | 13.26% | 307,146,368 | 11.22% | 743,125,454 | 17.32% | 904,769,393 | 28.93% | 391,997,649 | 16.88% |
| PiggyBac | 25,568,568 | 0.70% | 14,271,115 | 0.52% | 13,683,942 | 0.32% | 6,242,598 | 0.20% | 9,228,673 | 0.40% |
| Tourist/Harbinger | 3,509,956 | 0.10% | 6,656,014 | 0.24% | 10,579,236 | 0.25% | 4,453,869 | 0.14% | 2,199,720 | 0.09% |
| Other (Mirage, P-element, Transib) | 46,791,717 | 1.27% | 5,008,403 | 0.18% | 6,326,642 | 0.15% | 3,570,992 | 0.11% | 993,579 | 0.04% |
| Rolling-circles | 29,521,374 | 0.80% | 105,886,680 | 3.87% | 324,281,627 | 7.56% | 139,155,718 | 4.45% | 5,674,177 | 0.24% |
| Total interspersed repeats | 2,458,966,257 | 66.90% | 1,591,021,662 | 58.09% | 2,589,727,245 | 60.36% | 1,981,438,372 | 63.35% | 1,343,923,274 | 57.87% |

**Table S7.** Statistics of RNAseq mapping and assembly.

| Species | Samples | Total RNAseq reads number | HiSAT2 mapping rate % | StringTie mRNA number | StringTie exon number |
|---|---|---|---|---|---|
| *Mantis religiosa* | Sample_1 | 88,785,004 | 91.45% | 242,954 | 614,681 |
| | Sample_2 | 86,154,304 | 91.46% | | |
| | Sample_3 | 82,340,876 | 92.14% | | |
| | Sample_4 | 87,545,542 | 92.31% | | |
| | Sample_5 | 64,013,662 | 88.66% | | |
| | Sample_6 | 79,818,902 | 88.58% | | |
| | Sample_7 | 119,602,504 | 89.11% | | |
| *Tenodera sinensis* | Sample_1 | 67,885,312 | 94.91% | 24,901 | 180,176 |
| | Sample_2 | 74,463,660 | 94.68% | | |
| | Sample_3 | 69,051,400 | 94.34% | | |
| | Sample_4 | 73,588,644 | 93.78% | | |
| | Sample_5 | 70,815,820 | 94.53% | | |
| | Sample_6 | 65,008,126 | 93.26% | | |
| | Sample_7 | 67,108,790 | 93.57% | | |
| *Deroplatys truncata* | Sample_1 | 96,379,192 | 97.48% | 22,917 | 562,724 |
| | Sample_2 | 76,125,082 | 97.21% | | |
| | Sample_3 | 108,158,576 | 96.86% | | |
| | Sample_4 | 86,689,652 | 97.34% | | |
| | Sample_5 | 83,429,828 | 97.33% | | |
| | Sample_6 | 80,351,926 | 97.16% | | |
| | Sample_7 | 93,349,968 | 97.03% | | |
| *Hymenopus coronatus* | Sample_1 | 121,597,780 | 96.88% | 115,759 | 400,507 |
| | Sample_2 | 81,361,518 | 96.95% | | |
| | Sample_3 | 99,558,770 | 97.35% | | |
| | Sample_4 | 79,018,200 | 96.95% | | |
| | Sample_5 | 85,621,424 | 97.14% | | |
| | Sample_6 | 78,941,540 | 96.95% | | |
| | Sample_7 | 84,850,904 | 97.04% | | |

Note: The StringTie mRNA coordinates were converted into hints gff format, and used as the hints data in augustus gene prediction.

**Table S8.** Public genome data information.

| Species Name | Ploidy | Sequencing technology | Genome Size (bp) | Contig N50 size | Assembly level / Scaffold N50 size | Data Address | Data version |
|---|---|---|---|---|---|---|---|
| *Zootermopsis nevadensis* | 2n = 52 | Illumina HiSeq | 485,009,472 | 22.8 Kb | Scaffold level, 751.1 Kb | NCBI | ZooNev1.0 |
| *Blattella germanica* | 2n = 24 | Illumina | 2,037297555 | 12.1 Kb | Scaffold level, 1.1 Mb | NCBI | Bger_1.1 |
| *Deroplatys lobata* | 2n = 28 | nanopore 59.86 | 3,962,954,272 | 6.11 Mb | chromosome level, 285.05 Mb | CNCB | Dlob_genome_v1.0 |